\documentclass[11pt, a4]{article}
\usepackage[utf8]{inputenc}
\usepackage{physics,amsmath}
\usepackage{amsmath}
\usepackage{amssymb}
\usepackage{graphicx}
\usepackage[top=1in]{geometry}
\usepackage{float}
\usepackage{setspace}
\usepackage{titlesec}
\usepackage[hidelinks,citecolor=blue,urlcolor=blue,colorlinks=true,bookmarks=false,hypertexnames=true]{hyperref} 
\usepackage{natbib} 
\setcitestyle{numbers,square} 
\usepackage{wrapfig}
\usepackage{tikz}
\usetikzlibrary{positioning}
\usepackage{braket}
\usepackage{}
\usepackage[british]{datetime2}
\usepackage[page,toc,titletoc,title]{appendix}
\usepackage{tocloft}
\usepackage{blindtext}
\usepackage{dsfont}
\usepackage{dirtytalk}

\newcommand\numberthis{\addtocounter{equation}{1}\tag{\theequation}}
\usepackage{comment}
\usepackage{bigints}
\usepackage{caption}
\usepackage{xcolor}

\newcommand{\LmInertial}{\mathcal{L}^{\text{M}}_{\text{inertial}}}
\newcommand{\LmRindler}{\mathcal{L}^{\text{M}}_{\text{Rindler}}}
\newcommand{\LmRRindler}{\mathcal{L}^{\text{M}}_{\text{RRindler}}}
\newcommand{\Prf}{\mathcal{P}_r}
\newcommand{\Pone}{\mathcal{P}_1}
\newcommand{\Ptwo}{\mathcal{P}_2}
\newcommand{\De}{\Delta \mathcal{E}}

\newcommand{\Wp}{W_{+}}
\newcommand{\Wm}{W_{-}}
\newcommand{\Lp}{L_{+}}
\newcommand{\Lm}{L_{-}}

\DeclareMathOperator{\sign}{sign}

\begin{document}

\title{Quantum Field Theory in Successive Rindler Spacetimes
\vspace{-0.25cm} 
\author{Nitesh K. Dubey$ ^{a,b}$, Jaswanth Uppalapati$^{c}$, Sanved Kolekar$^{a,b}$,  \\ \vspace{-0.5cm} \\ 
\textit{$ ^a $Indian Institute of Astrophysics} \\ 
\textit{Block 2, 100 Feet Road, Koramangala,} 
\textit{Bengaluru 560034, India.} \\ 
\textit{$ ^b $Pondicherry University} \\ 
\textit{R.V.Nagar, Kalapet, Puducherry-605014, India}\\
\textit{$ ^c $Department of Physics and Astronomy, Rutgers University,} \\ 
\textit{126 Frelinghuysen Rd., Piscataway
NJ 08854, USA}\\
\texttt{\small Email: \href{mailto:nitesh.dubey@iiap.res.in}{nitesh.dubey@iiap.res.in}, 
\href{mailto:sanved.kolekar@iiap.res.in}{sanved.kolekar@iiap.res.in},\href{mailto:nu75@physics.rutgers.edu}{nu75@physics.rutgers.edu}}}}

\maketitle
\abstract{We study successive Rindler-like transformations in Minkowski spacetime and the corresponding sequence of vacuum states perceived by observers restricted to respective wedges. Interestingly, once more than one transformation is applied, the resulting multi-Rindler observers become non-uniformly accelerating. The characteristic trajectories, confined to nested wedges, exhibit characteristic accelerations and horizon shifts depending on transformation parameters ${g_1, g_2, \ldots, g_{n}}$. For the second-level transformation (\emph{Rindler Rindler} case), the late time acceleration asymptotically approaches $2g_2$ for one branch and diverges for the other. We study Minkowski, Rindler, and Rindler Rindler vacuum states from the perspective of Unruh–DeWitt (UDW) detectors along inertial, Rindler, and Rindler Rindler trajectories. The response of the UDW detector coupled to a real massless scalar field confirms the thermality: the transition rate of Rindler Rindler observer in Minkowski vacuum matches that of a standard Rindler detector with acceleration $2g_2$, yielding a Planckian spectrum at late times. The conclusions are discussed.}

\pagebreak

\tableofcontents

\pagebreak

\section{Introduction}
The observer dependence of the notion of particles in quantum field theory 
has long been recognised as a fundamental aspect of the theory of relativity. 
A particularly striking manifestation of this feature is the \emph{Unruh effect}, 
according to which a uniformly accelerated observer in Minkowski spacetime 
perceives the inertial vacuum as a thermal bath with temperature 
$T = \hbar a / 2\pi c k_B$, where $a$ is the magnitude of the proper acceleration 
~\cite{PhysRevD.14.870, PhysRevD.7.2850, RevModPhys.80.787}. This phenomenon demonstrates that the concept of vacuum depends on the observer's frame of reference and that thermal effects can emerge solely from acceleration and horizon structure, without invoking gravitation. In the Rindler description of Minkowski spacetime, the right (or left) Rindler wedge 
is causally disconnected from its complement, and the reduced density matrix, corresponding to the Minkowski vacuum, obtained by tracing over the inaccessible modes, corresponds to a thermal state. This construction provides a simple and precise connection between acceleration, causal horizons, and thermality, and is conceptually parallel to Hawking radiation in black hole spacetimes~\cite{Hawking:1975vcx, padmanabhan}. The Unruh effect has since been widely studied through different approaches, including detector models, quantum information perspectives, and curved-space generalizations~\cite{RevModPhys.80.787, Socolovsky:2013rga}. 

Given the thermal nature of the inertial (Minkowski) vacuum when described in Rindler coordinates, it is natural to ask whether there exist other classes of accelerated observers for whom the \emph{Rindler} vacuum itself appears thermally populated.
More generally, one may ask whether it is possible to define a hierarchy of observers, each obtained by applying a Rindler-like transformation to the previous one, such that the vacuum state of the $(n\!-\!1)$th observer is perceived as a thermal state by the $n$th.
This question was first addressed in the context of the \emph{Rindler--Rindler spacetime}, obtained by performing a second Rindler-like transformation within the Rindler wedge itself~\cite{Kolekar:2013hra}.
It was shown that the vacuum state defined by Rindler observers appears thermal to the Rindler--Rindler observers, in close analogy with how the Minkowski vacuum appears thermal to Rindler observers.
Furthermore, \cite{Kolekar:2013hra} found that a Rindler--Rindler observer in the Minkowski vacuum can be used as a proxy for certain aspects of the response of a uniformly accelerated (Rindler) observer immersed in a thermal bath of inertial quanta.

A related nested structure can also arise from successive deformations of null coordinates in near-horizon settings.
For instance, one may deform the Eddington--Finkelstein null coordinates as \( u \rightarrow - C\,U^{p} \) and \( v \rightarrow C\,V^{p} \), with \( p = a/\alpha \) and \(\{U,V\}\) denoting Kruskal null coordinates; such power-law reparametrizations can be viewed as compositions of Rindler-type maps acting on null coordinates.
A Planck-scale modifications of the location of an effective Rindler (or black-hole) horizon can lead to detector responses that display approximately thermal features in appropriate regimes \cite{Lochan:2021pio}, and the corresponding changes in an Unruh--DeWitt detector's excitation probability were analyzed in \cite{PhysRevD.111.045023}.
In models in which near-horizon physics is supplemented by a nonvanishing spacetime coordinate commutator, spatial and temporal shifts can become correlated, so that a small ``horizon displacement'' is more naturally interpreted as a dynamical near-horizon distortion rather than a purely static boundary shift.
Such distortions can modify the entanglement structure across the relevant null surface, the ingredient underlying the emergence of thermality upon restriction, and can also support transient, non-equilibrium contributions to fluxes and detector rates.
Within this qualitative picture, the Rindler-Rindler transformation provides a convenient analytic parametrization of time-dependent horizon deformations that relax to an asymptotic constant.
These considerations underscore that null hypersurfaces partition spacetime into causally disconnected regions, and that both vacuum assignment and detector response can be highly sensitive to the detailed horizon structure.

The iterative structure of successive Rindler transformations admits a natural interpretation in terms of progressively restricted observer algebras.
At each stage of the construction, the corresponding accelerated observer has access only to a subregion of the previously accessible spacetime, so that additional field degrees of freedom become causally inaccessible behind nested horizons.
Operationally, the state relevant to the $n$th observer can be viewed as obtained by restricting the global state to the algebra of observables accessible at that stage (equivalently, by tracing over an enlarged set of inaccessible modes), yielding a mixed state which, in the stationary cases or suitable late-time/adiabatic limits, exhibits thermal properties analogous to the standard Unruh construction.
From this perspective, the hierarchy of successive Rindler vacua may be viewed as defining a sequence of observer-dependent effective descriptions of the underlying quantum field theory.
Although the construction does not constitute a Wilsonian renormalization-group flow in the conventional sense, it exhibits several structurally similar features: namely, a hierarchical reorganization of accessible degrees of freedom, an emergent thermal characterization associated with reduced observables, and a recursive relation between descriptions adapted to different causal wedges.

This interpretation suggests possible connections with the algebraic formulation of quantum field theory, where observables restricted to causally accessible regions (such as horizons or wedges) are naturally described by local von Neumann algebras, whose associated modular automorphism groups encode the intrinsic dynamical/thermal structure seen by the restricted observer.
In the ordinary Rindler case, the Bisognano--Wichmann theorem identifies Lorentz boosts with the modular flow of wedge-localized algebras.
The successive Rindler construction raises the possibility that an analogous (though generally more intricate and potentially non-stationary) modular structure may underlie the recursive thermal features observed in such nested frames.
More broadly, the nested causal structure appearing in successive Rindler frames bears conceptual similarities to tensor-network and entanglement-based approaches to emergent geometry, in which spacetime organization is encoded in the pattern of accessible quantum correlations.
In this sense, repeated horizon restrictions may be interpreted as inducing an iterative reorganization of field entanglement, providing a potentially useful framework for investigating the interplay between causal structure, thermality, and observer-dependent notions of vacuum.

Another physically interesting aspect of the construction is that, for $n \geq 2$, the associated trajectories are typically non-stationary, with time-dependent accelerations and dynamically shifted horizons.
The resulting spacetimes therefore provide analytically tractable toy models for studying non-equilibrium horizon thermality and transient Unruh-like phenomena.
Since generic black-hole horizons in dynamical settings are likewise non-stationary, successive Rindler geometries may offer useful insight into horizon relaxation, evolving causal structure, and the role of entanglement in time-dependent near-horizon physics.

Previous works~\cite{Kolekar:2013hra, Lochan:2021pio, PhysRevD.111.045023} have examined Bogoliubov transformations arising from successive Rindler-like mappings or have computed detector responses for shifted Rindler trajectories in different vacua. However, it is important to note that under successive Rindler-like transformations, the time coordinate $t_n$ does \emph{not} correspond to the proper time along the Rindler trajectory in the $n$th wedge for $n \ge 2$. In the Bogoliubov analysis, it is the relativity of positive-frequency modes defined with respect to the $t_n$ and $t_{n-1}$ coordinates that generates the associated particle content. This naturally raises the question: who are the observers for whom $t_n$ is the proper time in the $n$th Rindler spacetime? Owing to the conformal nature of the transformations, one may already anticipate that these observers follow non-accelerating trajectories. Since detector responses need not, in general, agree with particle content inferred from Bogoliubov coefficients~\cite{sriramkumar_padmanabhan_2002, Padmanabhan:2019art}, this further motivates an investigation of the transition rates associated with these trajectories. In this work, we focus on these multi-Rindler trajectories and study the corresponding detector response in various vacua.

The paper is structured as follows. In Section~\ref {sec:restowedge}, we revisit the construction of an arbitrary number of successive Rindler-like transformations, and generate a hierarchy of \emph{multi-Rindler} observers labelled by an integer $n$. 
Each such observer, which interestingly is non-uniformly accelerating for $n \ge 2$, defines a distinct notion of vacuum and horizon, 
and the relations among them provide a rich framework for exploring 
the interplay between acceleration, causality, and thermality in 
flat spacetime. To probe what is perceived by these observers, we employ both a field-theoretic and a detector-based approach. Using Bogoliubov transformations in Section~\ref{sec:bogoltrans}, we reaffirm that the vacuum of the $(n-1)$th Rindler observer appears as a thermal state 
to the $n$th Rindler observer, consistent with the hierarchical 
structure of successive accelerations as shown in \cite{Kolekar:2013hra}. In Section \ref{sec4} we explicitly construct the non-uniformly accelerating trajectories associated with these observers, derive their proper accelerations, and analyse the corresponding causal structure and horizon shifts. Complementarily, in Section~\ref{sec:detector}, we analyze the transition probability of a Unruh--DeWitt (UDW) detector coupled to a massless scalar field in different vacuum states. 
For a detector following the Rindler--Rindler trajectory in the Minkowski vacuum, the transition probability, evaluated within the saddle-point approximation and in the corresponding asymptotic regime, coincides with that of a standard Rindler detector with acceleration $2g_2$, confirming the effective doubling of acceleration in this frame. We discuss various limits in which different transition probabilities reduce to each other. In particular, in the limit $g_1 \to 0$, an inertial observer in the Rindler--Rindler vacuum perceives physics identical to that of an inertial observer in the Rindler vacuum far from the horizon. In Section \ref{sec:31trans}, we show that at late times, the transition rate of the UDW detector along a Rindler Rindler trajectory in Minkowski vacuum becomes Planckian at temperature $2g_2/2\pi$, as expected, since the proper acceleration asymptotically reaches a constant value. This is interestingly different than the case of the non-Planckian result for the expectation value of the number operator obtained in Eq.(28) of \cite{Kolekar:2013hra} for a Rindler Rindler observer in Minkowski vacuum using the Bogoliuobov transformation at $t_0 = 0$ Cauchy slice. The result in Eq.(28) of \cite{Kolekar:2013hra} depends on both parameters $g$ and $g'$, while the late time response we obtain is dependent on $g'$ only, in particular, it is thermal with temperature $2g'/2\pi$. We conclude in Section~\ref{sec:conclusion} with a discussion of our findings 
and their implications for understanding thermal perception in 
successively accelerated frames. We use units where  $\hbar = c = k_B = 1$.

\section{Restricting QFT to a Wedge} \label{sec:restowedge}
In Minkowski spacetime, wedges are regions bounded by two non-parallel characteristic hyperplanes. They play an important role in many areas, including chiral conformal field theory, Wigner’s classification of elementary particles, and the near-horizon geometry of local horizons. In this section, we revisit the well-known Rindler wedge and subsequently define its generalisation.
\subsection{$n^{th}$ Rindler transformation}
One noteworthy restriction to the Minkowski space-time is R:= $\{ x \in \mathbb{R}^{1,3} | x^1 > |t|  \}$, called the right Rindler wedge. This wedge has a non-void commutant, called the left Rindler wedge. So, the modular group, which is generated by Lorentz boosts, is defined. The one-parameter group of Lorentz boost isometries provides a way to construct the Rindler spacetime. An observer travelling along the Lorentz boost  isometries in Minkowski spacetime can be described by the trajectory
\begin{equation} \label{eq:1}
    t_0 = \frac{e^{g_1 x_1}}{g_1} \sinh\bigl(g_1 \tau_1 e^{-g_1 x_1}\bigr), \qquad
    x_0 = \frac{e^{g_1 x_1}}{g_1} \cosh\bigl(g_1 \tau_1 e^{-g_1 x_1}\bigr),
\end{equation}
where $\tau_1$ is the proper time of the accelerated observer, and $x_1$ is a spacelike coordinate defined such that $g_1 e^{-g_1 x_1}$ corresponds to the observer's proper acceleration. The observer experiences the Minkowski vacuum as a thermal bath at a temperature $g_1 e^{-g_1 x_1}/(2\pi)$ \cite{PhysRevD.14.870}. 

By defining a timelike coordinate $t_1 = \tau_1 e^{-g_1 x_1}$, the pair $\{t_1, x_1\}$ forms a coordinate system, spanning the right Rindler wedge, known as the Rindler coordinates.
One can construct another spacetime, called the Rindler-Rindler spacetime \cite{Kolekar:2013hra}, by making another Rindler transformation to get a new coordinate $\{t_2,x_2\}$ as follows:
\begin{align}
    t_0 &= \frac{e^{g_1 x_1}}{g_1} \sinh{g_1 t_1}, &\! x_0 &= \frac{e^{g_1 x_1}}{g_1} \cosh{g_1 t_1}; \label{eq:1a} \\
    t_1 &= \frac{e^{g_2 x_2}}{g_2} \sinh{g_2 t_2}, &\! x_1 &= \frac{e^{g_2 x_2}}{g_2} \cosh{g_2 t_2} .\label{eq:1b}
\end{align}
\cite{Kolekar:2013hra} found, using Fock-space calculations with the Bogoliubov transformation, that the vacuum of the observers travelling along Lorentz boost isometries, defined in Eq.\eqref{eq:1}, appears thermal at $t_0 =0$ to the Rindler--Rindler observer, defined by setting coordinate time $t_2 = \text{proper time } \tau$. Further, \cite{Kolekar:2013hra} illustrated that the Minkowski vacuum appears to the Rindler-Rindler observer to be similar to what a Rindler observer sees in a thermal bath of Unruh-Minkowski particles. It turns out that one can generalize the transformation shown in the above equation, Eqs.\eqref{eq:1a}-\eqref{eq:1b}, as follows:
\begin{align*}
    t_0 &= \frac{e^{g_1 x_1}}{g_1} \sinh{g_1 t_1}, &\! x_0 &= \frac{e^{g_1 x_1}}{g_1} \cosh{g_1 t_1}; \\
    t_1 &= \frac{e^{g_2 x_2}}{g_2} \sinh{g_2 t_2}, &\! x_1 &= \frac{e^{g_2 x_2}}{g_2} \cosh{g_2 t_2}; \\
    &..............\\
    t_{n-1} &= \frac{e^{g_n x_n}}{g_n} \sinh{g_n t_n}, &\! x_{n-1} &= \frac{e^{g_n x_n}}{g_n} \cosh{g_n t_n}.  \numberthis \label{eq:2}
\end{align*}

The trajectory whose proper time $\tau$ equals the coordinate time corresponding to the $n^{\text{th}}$ Rindler transformation shown above is restricted to a region in Minkowski spacetime, called the shifted Rindler wedge\cite{Lochan:2021pio}, defined by
\begin{equation} \label{eq:3}
 S_n := \bigg\{ x \in \mathbb{R}^{1,3} \mid \bigg( x_0 - \frac{1}{g_1}  
    \exp\left(\frac{g_1}{g_2} \exp\Big(\frac{g_2}{g_3} 
    \exp\big(\tfrac{g_3}{g_4} \exp(\tfrac{g_4}{g_5} \exp(\cdots \exp(\tfrac{g_{n-2}}{g_{n-1}})))\big)\Big)\right) \bigg) > |t_0| \bigg\}.  
\end{equation}
In other words, Eq.\eqref{eq:3} describes the region of spacetime accessible causally to an observer traveling along the trajectory with proper time \(\tau = t_n  \). In this paper, we focus mainly on the special case $n=2$, known as the Rindler–Rindler transformation \cite{Kolekar:2013hra}. The corresponding space-time is designated as the right Rindler-Rindler wedge, defined by $S_2 :=\{ x \in \mathbb{R}^{1,3} \mid ( x_0 - 1/g_1  ) > |t_0| \}$. Just like the Rindler trajectory can be thought of as a consequence of a sequence of Lorentz boosts, the right Rindler-Rindler trajectories can be interpreted as a sequence of shifted Rindler trajectories at each time slice \cite{PhysRevD.111.045023, Barman:2025aqt}. 

\subsubsection{ Left RR wedge --- Commutant and analytical extension}
\begin{figure}[H]
    \centering
    \includegraphics[width=.92\textwidth]{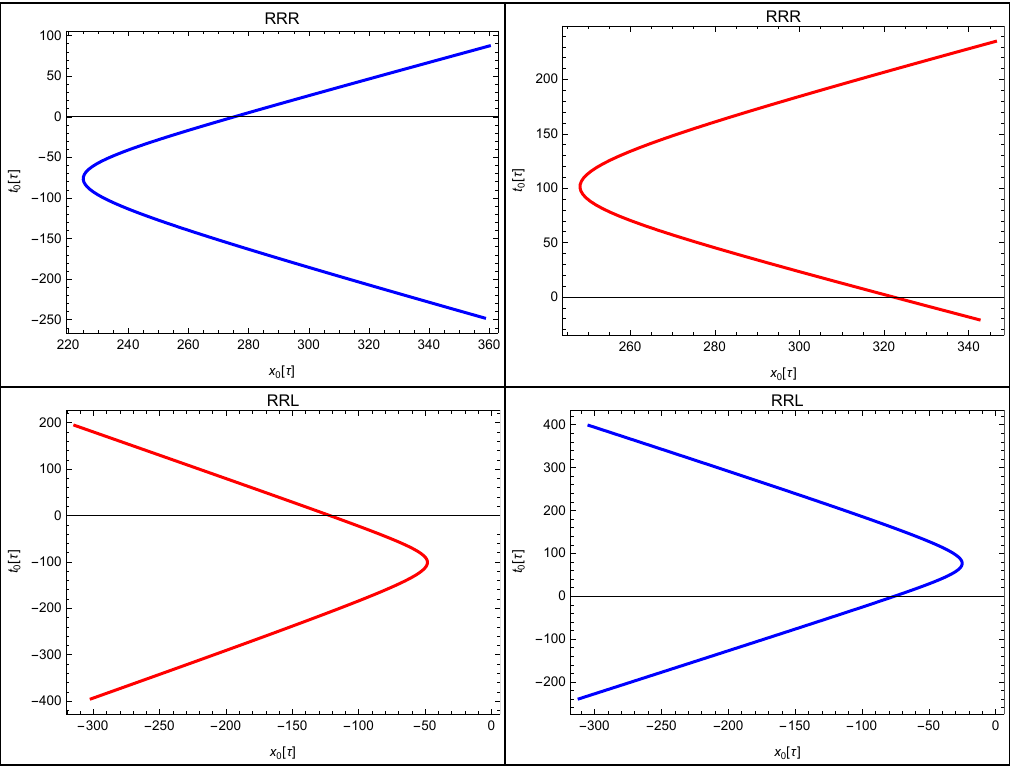}
     \captionsetup{margin=1cm, font=small}
    \caption{The plots above illustrate the Rindler--Rindler trajectories in the Minkowski plane, as introduced in Eq.~\eqref{eq:10} and further discussed in Section \ref{sec:numtraj}. The red curves correspond to the positive root of $\dot{y}$, while the blue curves represent the negative root. The label RRR denotes trajectories in the right Rindler--Rindler wedge, and RRL denotes those in the left Rindler--Rindler wedge. The parameters used are $g = g' = 0.01$.} 
    \label{fig:2}
\end{figure}

The coordinates patch obtained by the transformation, defined in Eq.\eqref{eq:2}, covers only a part of Minkowski spacetime. In order to cover the patch of Minkowski spacetime, which is the commutant of observables localised in the patch described by Eq.\eqref{eq:3}, one needs an analytic continuation. For $n=2$, the commutant of right Rindler-Rindler wedge, called left Rindler–Rindler wedge (LRR), is defined as \( S'_{2} := \left\{ x \in \mathbb{R}^{1,3} \;\middle|\;  \left(x_{0} - 1/g_{1}\right) < -|t_{0}| \right\}. \)
Applying time reversal together with reflection about the vertical line 
$x_{0} = 1/g_{1}$, i.e., \( (t_{0},x_{0}) \;\mapsto\; (-t_{0}, 2/g_{1} - x_{0}), \) one obtains the following coordinate patch that covers $S'_{2}$:
\begin{equation} \label{eq:4}
    t_0 = - \frac{e^{g_1 x_1}}{g_1} \sinh{g_1 t_1} , x_0 =\frac{2}{g_1} - \frac{e^{g_1 x_1}}{g_1} \cosh{g_1 t_1}
\end{equation}
and
\begin{equation} \label{eq:5}
    t_1 = \frac{e^{g_2 x_2}}{g_2} \sinh{g_2 t_2} , x_1 =  \frac{e^{g_2 x_2}}{g_2} \cosh{g_2 t_2}. 
\end{equation}
\\
The trajectory of an observer in the left Rindler-Rindler wedge, which is a mirror image of the trajectory with proper time equal to the coordinate time of the right Rindler-Rindler (RRR), and defined for all time, is illustrated in Fig.\ref{fig:2}. 

The dual of the right Rindler-Rindler wedge, defined above in Eqs.\eqref{eq:4}-\eqref{eq:5}, namely the left Rindler Rindler spacetime (LRR), is also a wedge. However, if the second transformation is defined with the minus sign as
\begin{equation} \label{eq:5b}
    t_1 = - \frac{e^{g_2 x_2}}{g_2} \sinh(g_2 t_2), \qquad 
    x_1 = - \frac{e^{g_2 x_2}}{g_2} \cosh(g_2 t_2),
\end{equation}
the quantum field theory becomes restricted to a diamond-shaped region within the full LRR wedge. The size of the diamond can be controlled by $g_1$ and $g_2$, and it can also be conformally mapped to a wedge without altering the causal structure of the left Rindler-Rindler patch. Since the massless scalar field in 1+1 dimensions is conformally invariant, the mapping of the diamond region to a wedge region by a suitable conformal transformation can be used to transfer the modular flow on the diamond to the wedge. Due to the finite lifetime of the diamond-shaped region, only $|x| +|t| <l$ is in causal contact. One can refer \cite{Martinetti:2002sz} for the effect of the finite lifetime of an observer/detector. In particular, it's known that the diamonds with sufficiently large $l$ have temperatures equivalent to the Unruh temperature. However, smaller diamonds corresponding to a short lifetime observer possess relatively higher temperatures. To study the properties of the diamond, one can also put a detector along this trajectory with compact switching so that it's switched on within the diamond only. The dual corresponding to $n^{th}$ transformation can be defined similarly.

We defined above different restrictions in both the left and right Rindler wedges. Now, we take a massless real scalar field and study its properties corresponding to the different restrictions introduced above, called $ n^{th}$ right Rindler wedge. The discussion for the left wedge can follow similarly.

\section{Global Fock space relations} \label{sec:bogoltrans}

The Bogoliubov transformations relating Minkowski spacetime, Rindler spacetime, and the Rindler-Rindler spacetime are known in the literature \cite{Kolekar:2013hra}. In this section, we reaffirm the Bogoliubov transformation that relates the $(n-1)^{th} $ Rindler spacetime to $n^{th} $ Rindler spacetime. The metric in terms of coordinates defined in different wedges is given by
\begin{equation} \label{eq:metric}
    ds^2 = - dt_0^2 + dx_0^2 = \Omega_{n-1}^2(-dt_{n-1}^2 + dx_{n-1}^2) = \Omega_{n}^2(-dt_{n}^2 + dx_{n}^2).
\end{equation}
Here, $\Omega_{n-1}$ and $\Omega_{n}$ are spacetime-dependent conformal factors. In particular, $\Omega_1$ and $\Omega_2$ are given by $\exp(g x_1)$ and $\exp(g x_1 + g' x_2)$, respectively. The conformal structure of the metric shown above in Eq.\eqref{eq:metric}, expressed in different coordinate patches, ensures that a plane wave mode decomposition is possible, allowing the real massless scalar field to be written as a sum of plane wave mode solutions to the Klein--Gordon (KG) equation: 
\begin{align*} 
\hat{\phi}(t,x) &=  \int _{-\infty} ^{\infty} \frac{d k_{n-1}} {(2 \pi)^{1/2}\sqrt{2 |k_{n-1}|}} (\hat{a}_{k_{n-1}} e^{i(k_{n-1} x_{n-1} - |k_{n-1}| t_{n-1})}  + \hat{a}^\dag _{k_{n-1}} e^{i(-k_{n-1} x_{n-1} + |k_{n-1}| t_{n-1})} ) \\ &=\int _{-\infty} ^{\infty} \frac{d k_{n}} {(2 \pi)^{1/2}\sqrt{2 |k_{n}|}} (\hat{b}_{k_{n}} e^{i(k_{n} x_{n} - |k_{n}| t_{n})}  + \hat{b}^\dag _{k_{n}} e^{i(-k_{n} x_{n} + |k_{n}| t_{n})} ) \numberthis \label{eq:field}.
\end{align*}
Here, $k_{n-1}$, $k_{n}$ represent $(n-1)^{th}$ and $n^{th}$ Rindler frame modes, respectively. In the metric shown in Eq.\eqref{eq:metric}, corresponding to the transformation
\begin{equation} \label{eq:6}
    x_{n-1} = \frac{e^{g_n x_n}}{g_n}\cosh{g_n t_n} \quad ; t_{n-1}  = \frac{e^{g_n x_n}}{g_n}\sinh{g_n t_n},
\end{equation}
the future directed unit normal to the surface $t_{n-1}$= const, is $n^0$= $e^{-g_{n-1} x_{n-1} - g_n x_n }$. Therefore, the spatial metric determinant $\gamma$ satisfies \( n^0 \sqrt{\gamma} = 1 \), and hence, the scalar product in all spaces obtained from these transformations is precisely that of Minkowski space. Using the scalar product of plane wave modes, shown in Eq.\eqref{eq:field}, on $t_{n-1}$ = 0 slice, we get the following Bogoliubov coefficients (see \cite{Kolekar:2013hra, PhysRevD.111.065004, PhysRevA.107.L030203} ):
\begin{eqnarray} \label{eq:8}
    \alpha(k_{n-1},k_n) &=& \theta(k_{n-1}k_n) \sqrt{\frac{k_n}{k_{n-1}}} G(k_{n-1},k_n);  \beta(k_{n-1},k_n) = \theta(k_{n-1}k_n) \sqrt{\frac{k_n}{k_{n-1}}} G(-k_{n-1},k_n);   \\
    G(k_{n-1},k_n) &=& \frac{1}{2\pi g_n} \Gamma \bigg(-\frac{ik_n}{g_n} \bigg) \exp \bigg(i \frac{k_n}{g_n}\ln{\frac{|k_{n-1}|}{g_n}} + \sign{(k_{n-1})} \frac{\pi k_n}{2g_n}    \bigg)
\end{eqnarray}

The above expressions for the Bogoliubov coefficients 
\(\alpha_{k_{n-1},k_n}\) and \(\beta_{k_{n-1},k_n}\) are independent of 
\(g_{n-1}\). Since these Bogoliubov coefficients are obtained from the 
Klein--Gordon inner product, the invariance of this inner product ensures 
that \(|\alpha_{k_{n-1},k_n}|^2\) and \(|\beta_{k_{n-1},k_n}|^2\) are 
time-independent. Further, one can see from Eq.\eqref{eq:6} that
\begin{equation} \label{eq:7}
    t_{n-1}^2 -   x_{n-1}^2 = - \frac{e^{2 g_n x_n}}{g_n ^2}.
\end{equation}
At any given instant, the above Eq.~\eqref{eq:7} describes a segment of a hyperbola with \(x_{n-1} = t_{n-1}\) as an asymptote. For \(n > 1\), however, \(x_n\) varies with proper time, as shown in Section \ref{sec4}. Thus, for \(n>1\), the associated vacua are time dependent, similar to cosmological vacua, but the Bogoliubov coefficients, being global relations between two Fock bases, encode this time dependence purely as a phase. However, to obtain the reduced state corresponding to the \((n\!-\!1)^{\text{th}}\) transformation vacuum, as viewed from the \(n^{\text{th}}\) Rindler frame, we must trace over the degrees of freedom that are not in causal contact. We therefore focus on the \(t_{n-1} = 0\) slice, since any other constant \((n\!-\!1)^{\text{th}}\) Rindler coordinate time slice would inevitably include unobservable modes in either the past or future wedges. The expectation value of the number operator $\hat{b}_{k_n}^\dag \hat{b}_{k_n}$ in the conformal vacuum of $(n-1)^{th}$ Rindler, at \(t_{n-1} = 0\), is given by
\begin{equation} \label{eq:particlenum}
    \langle \mathcal{N}_{n n-1} (|k_n|) \rangle = \int d |k_{n-1}| | \beta_{k_{n-1},k_n}|^2  = \frac{\delta(0)}{e^{2 \pi |k_n| / g_n} - 1}.
\end{equation}
The above expression, Eq.~\eqref{eq:particlenum}, suggests that at $t_{n-1} = 0$ (and thus on the $t_n = 0$ hypersurface), the $(n-1)^{th}$ Rindler vacuum appears thermal to the  $n^{th}$ Rindler observer. It is important to note, however, that no unique inverse Bogoliubov transformation exists; \(n^{\text{th}}\) Rindler spacetime covers only a portion of the full spacetime. Nevertheless, following the approach in \cite{PhysRevD.111.065004}, which constructs a class of Minkowski states yielding the Rindler vacuum, one can obtain a class of states in the \((n-1)^{\text{th}}\) Rindler spacetime corresponding to the \(n^{\text{th}}\) Rindler vacuum.

\begin{figure}
    \centering
    \includegraphics[width=.92\textwidth]{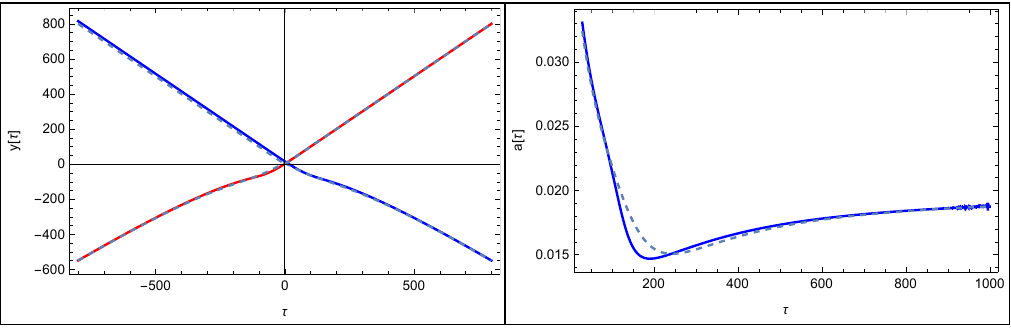}
     \captionsetup{margin=1cm, font=small}
    \caption{The left panel of the above plot depicts the trajectory $t_2 = \tau$ obtained from solving Eq.\eqref{eq:10} with $g=g'=0.01$, while the right panel shows the proper acceleration in the fourth quadrant of the left panel. Dashed lines indicate the approximate analytical solutions discussed in Sections \ref{subsec:negydot} and \ref{subsec:posydot} with $C=0$, while the solid curves correspond to the exact numerical solutions. The red colour represents the positive root of $\dot{y}$, while the blue colour represents the negative root. } 
    \label{fig:1}
\end{figure}

The relations discussed above are global in nature and do not necessarily correspond to what a localized observer in spacetime would measure \cite{sriramkumar_padmanabhan_2002}. A localized detector probes the spectral pattern of vacuum fluctuations, which includes contributions beyond particle-like excitations \cite{Padmanabhan:2019art}. Consequently, the detector’s response generally does not coincide with the results of the Bogoliubov coefficient calculation. Computing the detector’s response requires specifying the trajectory for which the proper time $\tau$ equals the coordinate time $t_n$. However, the coordinate time \(t_n\) used to define the positive- and negative-frequency modes in the field of Eq.~\eqref{eq:field} does not correspond to the proper time of an observer at \(x_n = \text{constant}\) for \(n > 1\). Therefore, in the next section, we determine the trajectory for which the coordinate time coincides with the observer’s proper time.

\section{Characteristic trajectory} \label{sec4}

Here, we discuss the characteristic trajectory along which the coordinate time $t_n$ matches the proper time of the observer. Since the metric resulting from more than one Rindler transformation lacks time-translation symmetry, observers at \(x_n = \text{constant}\) do not follow trajectories for which the proper time coincides with the coordinate time associated with the \(n^\text{th}\) transformation, for $n>1$. For $n=2$, the world line of such  an observer $x_2 \equiv y[\tau]$ is determined by the following line element:
\begin{equation} \label{eq:9}
    ds^2 =- d\tau ^2 = e^{2 g x_1 + 2 g' y } (- d\tau ^2 + dy^2),
\end{equation}
which boils down to solving the following nonlinear differential equation:
\begin{equation} \label{eq:10}
    \dot{y}^2 = 1 - e^{-2 \frac{g}{g'} e^{g' y} \cosh{g' \tau} - 2 g' y } .
\end{equation}
Here, in the above expression, we have renamed $g_1$ as $g$ and $g_2$ as $g'$, following the notation in \cite{Kolekar:2013hra}, since we will be restricted to $n = 2$ for most of the discussion. We do not have an exact analytical solution to the above equation, Eq.~\eqref{eq:10}. However, we can solve it numerically or analytically with some approximations described in the subsections below.

\subsection{Asymptotic analytical solution}

The differential equation \eqref{eq:10} can be rewritten as
\begin{align} 
    \dot{y} =& \pm \sqrt{ 1 - \exp{-2 \frac{g}{g'} e^{g' y} \cosh{g' \tau} - 2 g' y }} \numberthis \label{eq:11}\\
    \approx & \pm \bigg(  1 - \frac{1}{2} \exp{-2 \frac{g}{g'} e^{g' y} \cosh{g' \tau} - 2 g' y } + ... \bigg).  \numberthis \label{eq:12}
\end{align}
In the second line, the binomial expansion is applied. For $\tau \rightarrow \infty$ and $g' \neq 0$, one can ignore the higher order terms, keeping only the first two terms of the expansion, since $\exp{-2 g \exp{g'(y+\tau)}/g'-2g'y} << 1$. We now proceed to solve the different cases below.

\subsubsection{Late time behavior, negative $\dot{y}$} \label{subsec:negydot}
In this subsubsection, we discuss the negative root of Eq.~\eqref{eq:12} in the limit $\tau \rightarrow \infty$ with $g' \neq 0$. Performing a change of variable, $v = y + \tau$, the negative root branch of Eq.~\eqref{eq:12}, keeping only the first two terms, can be rewritten as
\begin{equation} \label{eq:13}
    2 \exp{ \frac{g}{g'} e^{g'v} + 2 g' v} dv = e^{2 g' \tau} d \tau .
\end{equation}
Performing the integration of both sides, one gets
\begin{equation} \label{eq:14}
     \frac{g'}{g^2} \exp{\frac{g}{g'} e^{g' v}} \bigg( \frac{g}{g'} e^{g' v} -1 \bigg)  =\frac{1}{4 g'} e^{2 g' \tau} + c,
\end{equation}
which can be inverted to get the following expression for $y$:
\begin{equation} \label{eq:15}
        y(\tau) = -\tau + \frac{1}{g'}\log(\frac{g'}{g}\bigg( 1+W \bigg( \frac{g^2e^{2g'\tau-1}}{4g'^2} + C \bigg) \bigg) ). 
\end{equation}
Here, \(W(\cdot)\) denotes the Lambert $W$ function. The integration constant $C$ can be set to be 0 by an appropriate choice of the boundary condition. For illustration, we show in Fig.\ref{fig:1} the above approximate solution, which closely matches the exact numerical solution at large proper times. The above solution, in Eq.\eqref{eq:15}, can be used to get the following shift of the horizon observed by the Rindler-Rindler observer at late times in the Minkowski spacetime inertial coordinates:
\begin{equation} \label{eq:16}
    x_0 - t_0 \rightarrow \frac{1}{g}.
\end{equation}
The above equation~\eqref{eq:16} implies that, at late times, the trajectory asymptotically approaches \( x_0 -t_0 = 1/g \), which corresponds to the shift of the Rindler--Rindler wedge shown in Eq.~\eqref{eq:3}. Furthermore, the proper acceleration at late times, obtained from this solution, is  
\begin{equation} \label{eq:17}
    a(\tau) = 2g' - \frac{2g'}{W\!\left( \frac{g^2 e^{2g'\tau-1}}{4g'^2} \right)} - \ldots ,
\end{equation}
which asymptotically approaches \( 2g' \) at late times, consistent with expectations for a shifted Rindler trajectory.

\subsubsection{Late time behavior, positive $\dot{y}$} \label{subsec:posydot}
Now we discuss the positive root of Eq.~\eqref{eq:12} in the limit $\tau \rightarrow \infty$ with $g' \neq 0$. Performing a change of variable, $v = \exp{g'(y + \tau)}$, the positive root branch of Eq.~\eqref{eq:12}, keeping only the first two terms, can be rewritten as
\begin{align} 
   \frac{1}{2 g' v} \bigg( 1- \frac{1}{4} e^{- gv/g'}  \bigg) ^{-1} dv = & d\tau \\
   \implies \frac{1}{2 g' v} \bigg( 1 + \frac{1}{4} e^{- gv/g'}  \bigg) dv \approx & d\tau \label{eq:19}
\end{align}
Here, once again, we have used the binomial expansion in the second line and kept only the first two terms, which is valid in the same approximation illustrated in the previous subsection. Performing the integration of both sides \cite{Gradshteyn2007}, one gets
\begin{equation} \label{eq:20}
     4 \log{v} + Ei(-gv/g') = 8 g'\tau + constant,
\end{equation}
which can be inverted using the Lagrange–Bürmann inversion method  (inversion using power series) to give
\begin{equation} \label{eq:21}
    y(\tau) = \tau + \sum_{k=1}^{\infty} 
    \frac{1}{k!}
    \left. 
    \frac{d^{\,k-1}}{dy^{\,k-1}}
    \left[
        \left(
            -\frac{1}{4g'} \,
            Ei\!\left(
                -\frac{g}{g'} e^{g'(y+\tau)}
            \right)
        \right)^{\!k}
    \right]
    \right|_{y=\tau}.
\end{equation}
Here, $Ei$ denotes the exponential integral function, and the boundary conditions are chosen such that the integration constant vanishes. The series in the above equation \eqref{eq:21} decays at a large time, acting as a correction to the $y=\tau$ line. This corresponds to the shift of horizon observed by the Rindler-Rindler, given by 
\begin{equation} \label{eq:16b}
    x_0 - t_0 \rightarrow \frac{1}{g} e^{g/g'},
\end{equation}
which is different from Eq.\eqref{eq:16} for the late time solution corresponding to the negative $\dot{y}$. However, since the exponential of a positive number is always greater than 1, the trajectory is still restricted to the Rindler-Rindler wedge $S_2$. Meanwhile, the acceleration corresponding to this solution, as shown in Appendix[\ref{Appendix B}], goes to infinity at late times. The early-time solution, that corresponds to the limit $\tau \rightarrow -\infty$, for negative $\dot{y}$ is the time reverse ($\tau \rightarrow -\tau$) of the solution shown in Eq.\eqref{eq:21}. Similarly, the early-time solution for the positive $\dot{y}$ case in the previous subsection is the time reverse of Eq.\eqref{eq:15}. 

In the next subsubsection, we discuss the characteristic trajectory, obtained numerically, without the assumptions used above to obtain the asymptotic analytical solutions.

\subsubsection{Numerical solution for interpolating regimes} \label{sec:numtraj}
The analytical solutions presented in the preceding subsubsections characterize the detector’s trajectory only in the early- and late-time regimes. However, the metric in Eq.~\eqref{eq:9} is not invariant under translations of the coordinate time \(t_2\), since \(x_2\), being a function of the Rindler-Rindler observer’s proper time, renders the conformal factor \(\exp(g x_1 + g' x_2)\) explicitly time dependent. Although the spacetime remains conformally flat, so that a massless scalar field in \(1+1\) dimensions can still be mapped to the Minkowski theory, the vector field \(\partial/\partial t_2\) is not a Killing vector. Consequently, the notion of positive frequency associated with \(t_2\)-evolution is nonstationary. Moreover, the acceleration discussed in the previous subsections is variable, which can render the quantum field dynamics non-Markovian. In such cases, memory effects arise, and the detector’s evolution can depend on its entire past history~\cite{Dubey:2025wws}. Therefore, to capture the full quantum field dynamics, one must determine the complete trajectory, including the interpolating regimes between early and late times. This motivates us to solve the nonlinear ordinary differential equation in Eq.~\eqref{eq:10} numerically, with the results shown in Fig.\ref{fig:1} and Fig.~\ref{fig:2}.

As expected, one can see from Fig.\ref{fig:1} and Fig.\ref{fig:2} that the characteristic trajectories are not hyperbolas in the Minkowski plane. This is also evident analytically, because $x_2$ in Eq.\eqref{eq:7} is a function of the proper time of the Rindler-Rindler observer. Furthermore, the trajectories in Fig.\ref{fig:1} and Fig.~\ref{fig:2} are not invariant under time reversal. This can also be understood analytically from the fact that $x_2 \equiv y[\tau]$ in Eqs.~\eqref{eq:15} and \eqref{eq:21} is not invariant under time reversal; consequently, the worldline in Eq.~\eqref{eq:7} is also not expected to be invariant under time reversal. However, one can see that the trajectories are restricted to a particular wedge, and the left and right Rindler-Rindler wedge trajectories are related by \( (t_{0},x_{0}) \;\mapsto\; (-t_{0}, 2/g - x_{0}) \). We also notice that the turning point in the trajectory is not $t_0 =0$, which is due to the chosen boundary conditions.

In the right panel of Fig.~\ref{fig:1}, we show the acceleration for $n=2$ with the negative $\dot{y}$ root, which can be seen to asymptotically approaching $2 g_{2}$. This is in agreement with what one expects from Eq.~\eqref{eq:17}. One can contrast this with the fact that the shift for the Rindler-Rindler trajectory horizon is $1/g_1$. The numerical calculations for $n=3$ and $n=4$ suggest that for the nth Rindler, with $n \geq 2$, the proper acceleration asymptotes to $2 g_n$ for an appropriate branch of solutions. Moreover, the shift of the horizon corresponds to\footnote{Note that here the assumption that $x_n<<t_n$ and $t_n \rightarrow \infty$ is implied for the horizon.} 
\begin{equation}
x_0-t_0 \approx \frac{1}{g_1}  
    \exp\left(\frac{g_1}{g_2} \exp\Big(\frac{g_2}{g_3} 
    \exp\big(\tfrac{g_3}{g_4} \exp(\tfrac{g_4}{g_5} \exp(\cdots \exp(\tfrac{g_{n-2}}{g_{n-1}})))\big)\Big)\right),
\end{equation}
which can be seen from Eq.\eqref{eqs:shift} in Appendix \ref{appenA}.

Having discussed the characteristic trajectory for the Rindler-Rindler wedge, in the next section, we introduce a two-level detector along this trajectory and study its transition probability and transition rate.

\section{UDW Detector response} \label{sec:detector}
A detector formalism is frequently introduced to determine the observer's perception. There are various theoretical models for detectors, such as Unruh-DeWitt (UDW) detectors, harmonic oscillators, atomic detectors, localised quantum fields, etc \cite{PhysRevD.101.045017, PhysRevD.103.025007, PhysRevD.109.045018, PhysRevD.105.065016, Ruep:2021fjh}. Among these, the UDW detector, a two-level system, is the simplest and most widely used, and we adopt it in our construction. The response of an amplitude-coupled UDW detector depends on the pullback of the Wightman two-point function along the trajectory of the detector. The Wightman two-point function of the real massless scalar field in 1+1 dimensions is infrared divergent. However, the momentum-coupled UDW detectors in 1+1 dimensions don't exhibit any infrared divergence. Furthermore, the ultraviolet properties of momentum-coupled UDW detectors in 1+1 dimensions are equivalent to the ultraviolet properties of the detector in 3+1 dimensions coupled to the amplitude of the field. This motivates to choose a UDW detector in 1+1 dimensions coupled with the momentum of the field.  

The detector is assumed to move along a timelike trajectory and interact locally with the quantum scalar field along its worldline.  Introducing the detector worldline \(x^\mu(\tau)\), parametrized by the proper time \(\tau\), the corresponding four-velocity is defined as
\begin{align}
u^\mu(\tau) \equiv \frac{dx^\mu(\tau)}{d\tau}.
\end{align}
We consider a derivative coupling model in which the detector couples to the variation of the field along the trajectory rather than directly to the field amplitude itself. The interaction action may therefore be written in the covariant form
\begin{equation}
\hat{S}_{\mathrm{int}}
=
\lambda
\int d\tau \,
\chi(\tau)\,
\hat{\mu}(\tau)\,
u^\mu(\tau)\nabla_\mu
\hat{\phi}\!\left(x(\tau)\right).
\label{eq:covariant-derivative-coupling-action}
\end{equation}
Here, \(\lambda\) is a small coupling constant controlling the strength of the interaction, while \(\hat{\mu}(\tau)\) denotes the monopole moment of the detector in the interaction picture. The function \(\chi(\tau)\) is a smooth switching function which regulates the duration of the interaction and ensures that the coupling is turned on and off smoothly in proper time. We choose a Gaussian switching function of the form $ \chi(\tau)=\exp(-(\tau- \tau_0)^2/\sigma^2)$ where \(\tau_0\) specifies the center of the interaction interval and \(\sigma\) determines the characteristic interaction timescale.

Since \(\hat{\phi}\) is a scalar field, its pullback along the detector trajectory satisfies
\begin{equation}
u^\mu(\tau)\nabla_\mu \hat{\phi}\!\left(x(\tau)\right)
=
\frac{d}{d\tau}\hat{\phi}\!\left(x(\tau)\right).
\end{equation}
Thus, the contraction of the field gradient with the detector four-velocity simply corresponds to the proper-time derivative of the field evaluated along the worldline. Using this relation, the interaction action can be rewritten in the simpler form
\begin{equation}
\hat{S}_{\mathrm{int}}
=
\lambda
\int d\tau \,
\chi(\tau)\,
\hat{\mu}(\tau)\,
\frac{d}{d\tau}
\hat{\phi}\!\left(x(\tau)\right).
\label{eq:worldline-derivative-coupling}
\end{equation}

The above expression makes explicit that the detector responds to the rate of change of the field along its trajectory. Choosing the detector proper time \(\tau\) as the evolution parameter, the corresponding interaction Hamiltonian is obtained as\cite{Juarez-Aubry:2014jba}
\begin{equation}
\label{eq:22}
\hat{H}_{\mathrm{int}}
=
\lambda
\chi(\tau)
\hat{\mu}(\tau)
\frac{d}{d\tau}
\hat{\phi}\!\left(x(\tau)\right).
\end{equation} 
In perturbation theory, the transition probability of the detector is therefore governed by two-point correlation functions of the differentiated field operator evaluated on the detector worldline. The transition probability of the detector, upto the leading order in the coupling constant, can be written as \cite{Birrell_Davies_1982}
\begin{equation}\label{eq:24a}
    \mathcal{L} = \lambda ^2 \int _{-\infty} ^{\infty} d\tau \int _{-\infty} ^{\infty} d\tau ^{'} \chi  (\tau) \chi (\tau ^{'} ) e^{-i \Omega (\tau- \tau ^{'})} \mathcal{A}^\alpha _{\text{tra}} (x(\tau),x(\tau ^{'})),
\end{equation}
where  $\mathcal{A}^\alpha _{\text{tra}} (\tau', \tau'')$ =  $\partial_\tau \partial_{\tau'} \mathcal{W}^\alpha _{\text{tra}} (\tau', \tau')$, and $\mathcal{W}^\alpha _{\text{tra}} (\tau', \tau')$ represent the pullback of the Wightman function along the trajectory of the detector. The superscript $\alpha$ specifies the chosen state and the subscript  \textit{tra} labels the trajectory, with $\mathcal{A}^\alpha_{\text{tra}}$ termed the momentum two-point function.

The metric corresponding to the \(n\)th Rindler transformation, given in Eq.~\eqref{eq:metric}, is conformally flat. As a result, the Wightman function acquires a particularly simple form due to the conformal invariance of a massless scalar field in \(1+1\) dimensions. To make this structure explicit, we introduce the null coordinates
\begin{equation}
u_n=t_n-x_n, \qquad v_n=t_n+x_n.
\end{equation}
For a massless scalar field, the Klein--Gordon equation is
\begin{equation}
\square \phi =0 .
\end{equation}
In \(1+1\) dimensions, the conformal factor drops out:
\begin{equation}
\square \phi
=
\frac{4}{\Omega_n^2}\partial_{u_n}\partial_{v_n}\phi ,
\end{equation}
so that the field equation reduces to
\begin{equation}
\partial_{u_n}\partial_{v_n}\phi =0 .
\end{equation}
Therefore, the mode decomposition and the corresponding conformal vacuum have the same functional form as in Minkowski spacetime, expressed in terms of the coordinates \((t_n,x_n)\). Therefore, the Wightman function for a massless scalar field in \(1+1\) dimensions corresponding to the \(n^{\text{th}}\) Rindler transformation, in the conformal vacuum state of the inertial observer, is given by
\begin{equation} \label{eq:25}
    \mathcal{W} (\textbf{x},\textbf{x}') = \frac{-1}{4 \pi} \ln{\left[\mu ((\Delta x_n)^2 - (\Delta t_n - i\epsilon)^2)\right]},
\end{equation}
where $\mu$ is an infrared cutoff, and \(\{x_n, t_n\}\) are defined in Eq.~\eqref{eq:2} for the \(n^{\text{th}}\) Rindler transformation.

\subsection{Minkowski vacuum} \label{tranprobMink}

Symmetries of the Minkowski spacetime allow us to define a 
global vacuum, invariant under Poincaré transformations, called the Minkowski vacuum. An eternal inertial detector with a sufficiently large energy gap, coupled to the field in the Minkowski vacuum state, is expected to observe a vanishing temperature. However, if one switches on a UDW detector with a finite energy gap along an inertial trajectory for a finite time, a nonzero transition probability can be obtained. Furthermore, a detector along an arbitrary trajectory can observe various excitations due to non-trivial Bogoliubov relations. In this subsection, other than the inertial trajectory, we also discuss and compare the response of the UDW along two other interesting trajectories, namely: (i) a uniformly accelerated trajectory, restricted to the right Rindler wedge, and (ii) the Rindler–Rindler trajectory, restricted to the shifted right Rindler wedge corresponding to the second Rindler transformation discussed in the previous section.

The Wightman function for a massless scalar field in 1+1 dimensions in the Minkowski vacuum state is given by 
\begin{equation} \label{eq:25a}
    \mathcal{W} (\textbf{x},\textbf{x}')  = \frac{-1}{4 \pi} \ln{[\mu ((\Delta x)^2 - (\Delta t - i\epsilon)^2})].
\end{equation}
For the detector along inertial trajectory $(x(\tau),t(\tau)) = (C,\tau)$, coupled to the scalar field momentum in the Minkowski vacuum state, the above Wightman function \eqref{eq:25a} gives the following two-point function relevant for the momentum coupling: 
\begin{equation} \label{eq:26a}
   \mathcal{A} ^{\text{M}}_{\text{inertial}} (\tau, \tau') = -\frac{1}{2\pi(\tau-\tau'-i\epsilon)^2}.
\end{equation}
Substituting the above two-point function Eq.\eqref{eq:26a} in Eq.\eqref{eq:24a}, and using the saddle point approximation to evaluate the integral (see Appendix[\ref{Appendix C}]), one gets the following expression for the transition probability
\begin{equation} \label{eq:27a}
    \LmInertial = \frac{\lambda^2 \, e^{-\frac{1}{2} \Omega^2 \sigma^2}}{2\, \Omega^2 \sigma^2} . 
\end{equation}
The above expression is independent of the proper time and the position $C$ of the inertial observer, as expected due to the Poincaré invariance of the Minkowskian vacuum. The limit $\Omega \sigma \rightarrow \infty$ can be consistently taken in Eq.~\eqref{eq:27a}. This is because the correlator $\mathcal{A} ^{\text{M}}_{\text{inertial}}$ contains only the usual short-distance pole near the real axis and lacks any imaginary-time periodic singularities. As a result, deforming the contour to the complex saddle point does not pick up additional contributions. The corresponding asymptotic behaviour, $L^{M}_{\mathrm{inertial}}\sim \sigma^2 e^{-\Omega^2\sigma^2/2}\to0$ therefore correctly reproduces the vanishing excitation probability of an eternally coupled inertial detector in the Minkowski vacuum.

\begin{figure} [H]
    \centering
    \includegraphics[width=.92\textwidth]{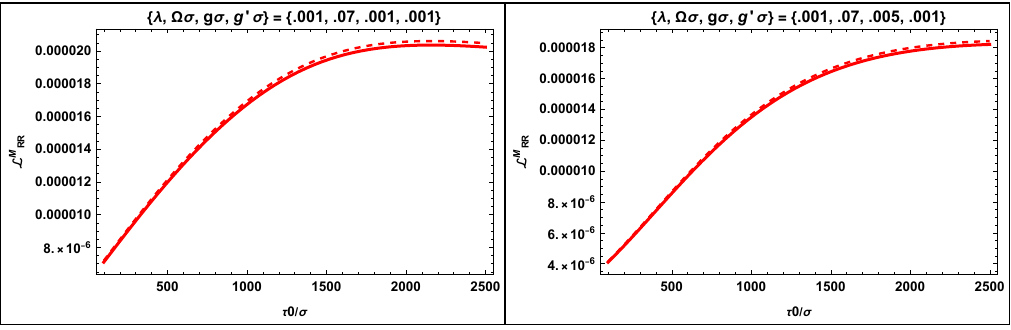}
     \captionsetup{margin=1cm, font=small}
    \caption{The dashed curves in the above plots illustrate the response of the UDW detector along the Rindler-Rindler trajectory in the Minkowski vacuum, computed using the saddle-point approximation, while the solid curves illustrate the results obtained through numerical computation.} 
    \label{fig:comparenumana}
\end{figure}

Now we choose the trajectory to be a one-parameter group of Lorentz boost isometries whose orbits can be understood as uniformly accelerated observers restricted to the Rindler wedge corresponding to ($t_1,x_1$), as described in Eq.~\eqref{eq:2}. For a Rindler observer, the trajectory can be written as
\begin{equation} \label{eternrintr}
    (x(\tau),t(\tau)) =   \bigg(\frac{1}{g}\cosh{(g \tau)},\frac{1}{g}\sinh{(g \tau)} \bigg) .
\end{equation}
The pullback of the momentum two-point function along the above trajectory, Eq.\eqref{eternrintr}, is
\begin{equation} \label{eq:Rindlermomntcorr}
    \mathcal{A} ^{\text{M}}_{\text{Rindler}} (\tau, \tau') = - \frac{g^2}{8 \pi \sinh^2{(g(\tau - \tau' - i \epsilon)/2)} }.
\end{equation}
Owing to the stationarity of the uniformly accelerated (Rindler) trajectory, the above two point function in the Minkowski vacuum, Eq.\eqref{eq:Rindlermomntcorr}, depends only on the proper time difference, thereby defining a stationary correlation function with equilibrium thermality. Furthermore, the above two-point function, Eq.\eqref{eq:Rindlermomntcorr}, is periodic in imaginary proper time and satisfies the Kubo--Martin--Schwinger (KMS) condition \cite{PhysRev.115.1342, PhysRevD.104.065001}.  Since the KMS property is fundamentally a statement about this pulled-back two-point function, an Unruh--DeWitt detector coupled through the interaction Hamiltonian in Eq.~\eqref{eq:22} and following the Rindler trajectory exhibits the corresponding thermal response. This thermality, however, should be understood at the level of the field correlation function and the asymptotic detector transition rate, rather than as a statement that the finite-time response is exactly Planckian for arbitrary switching profiles. In particular, for a detector switched on for a finite duration with a Gaussian switching function of width $\sigma$, the response function retains a nontrivial dependence on $\sigma$. Consequently, the satisfaction of the KMS condition alone does not guarantee an exactly Planckian spectrum for arbitrary finite interaction times or switching functions. Rather, the emergence of exact thermality depends nontrivially on the scaling of the interaction duration with the detector energy gap, as well as on the manner in which the switching profile is implemented. In particular, thermality is recovered in the appropriate long-interaction (or late-time) limit when the switch-on and switch-off intervals are stretched together with and the switching function possesses sufficiently rapid Fourier decay, with the relevant timescale set by the acceleration parameter $g$ \cite{Fewster:2016ewy}. Furthermore, in the limit $g \rightarrow 0$, the above two point function $\mathcal{A} ^{\text{M}}_{\text{Rindler}} $ approaches $ \mathcal{A} ^{\text{M}}_{\text{inertial}}$, the two point function for an inertial observer. Thus, as a consistency check, all properties that depend on the pullback of the two-point function along the detector's trajectories will be identical for both, inertial and Rindler, trajectories in the $  g\rightarrow 0$ limit.

The transition  probability, defined in Eq.\eqref{eq:24a}, using the saddle point approximation discussed in Appendix[\ref{Appendix C}] evaluates to
\begin{equation} \label{eq:MinkvacRind}
    \LmRindler = \frac{g^2 \sigma^2 \lambda^2}{8 \sin^2\left( \frac{g \Omega \sigma^2}{2} \right)} e^{-\frac{1}{2} \Omega^2 \sigma^2} + \text{residue contribution} ,
\end{equation}
which is again independent of the proper time of the Rindler observer. The factor \(e^{-\Omega^2\sigma^2/2}\) provides a Gaussian suppression for large detector energy gaps \(\Omega\), while the denominator \(\sin^2(g\Omega\sigma^2/2)\) encodes the non-trivial dependence on the acceleration \(g\). In $g \rightarrow 0 $ limit, the form of $ \LmRindler$ reduces to that of $\LmInertial$ in Eq.\eqref{eq:27a}. We further notice that if one increases the acceleration, keeping all other parameters fixed, the transition probability increases. This behavior is consistent with the fact that an accelerated observer perceives the Minkowski vacuum as populated with excitations, as expected from the Unruh effect. In the limit $\sigma \to \infty$, the first term in Eq.~\eqref{eq:MinkvacRind} vanishes. However, taking the limit $\sigma \to \infty$ causes an infinite number of poles to be crossed during the contour deformation. Consequently, the residue contributions dominate, causing the transition probability to diverge linearly. Nevertheless, one can still get the following finite rate (see Appendix [\ref{App:sumResidue}]):
\begin{equation} \label{eq:defineNewqqua}
\Gamma
=
\lim_{\sigma\to\infty}
\frac{\LmRindler}{\sqrt{2\pi}\sigma}
=
\frac{\lambda^2}{2}
\frac{\Omega}{
e^{2\pi\Omega/g}-1
}.
\end{equation} 
Interestingly, the above Eq.\eqref{eq:defineNewqqua} represents the Planckian transition rate with temperature associated with the Unruh effect. 
\begin{figure}[H]
    \centering
    \includegraphics[width=.92\textwidth]{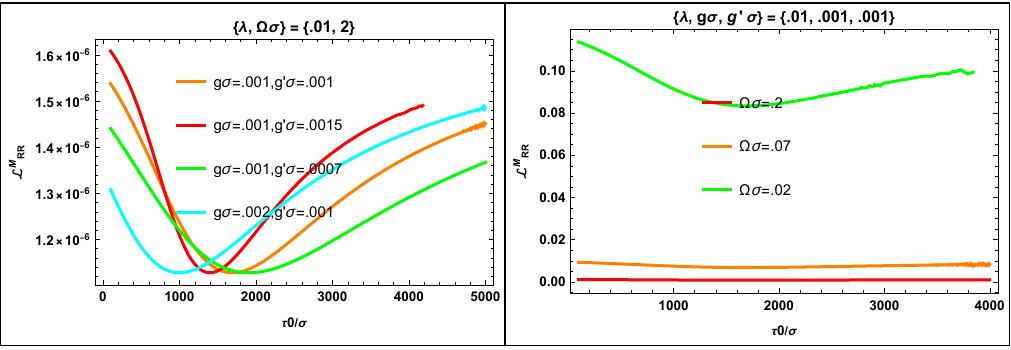}
     \captionsetup{margin=1cm, font=small}
    \caption{The above plots illustrate the transition probability of a UDW detector along the Rindler-Rindler trajectory, interacting with a real massless scalar field in the Minkowski vacuum, in 1+1 spacetime dimensions, as discussed in Section \ref{tranprobMink}. The red curve in the left panel is truncated near $\tau_0/\sigma=4000$ due to numerical limitations.} 
    \label{fig:9}
\end{figure}

Denoting \( \zeta = g^2/4g'^2 \), 
\( W_{\pm} =  W \left( \zeta \, \exp{2g'(\tau_0 \pm i\Omega\sigma^2/2) - 1} \right) \), and 
\( L_{\pm} = \tau_0 \pm i\Omega \sigma^2/2 \), the transition probability of the UDW detector along the Rindler-Rindler trajectory in the Minkowski vacuum, using the saddle point approximation with vanishing residue contribution, discussed in Appendix[\ref{Appendix C}], can be written as 

\begin{align*}
\LmRRindler &= -\sigma^2 \lambda^2 e^{-\frac{1}{2} \Omega^2 \sigma^2} \frac{(\Pone + \Ptwo)\, \Prf}{ (\De)^2} \numberthis
\end{align*}
where,
\begin{align*}
\Prf 
&= \frac{g'^2}{(1 + \Wp)(1 + \Wm)} , \\
\De 
&= e^{(1 + \Wp) e^{-2g' \Lp} + 1 + \Wp} - e^{(1 + \Wm) e^{-2g' \Lm} + 1 + \Wp}  - e^{(1 + \Wp) e^{-2g' \Lp} + 1 + \Wm} + e^{(1 + \Wm) e^{-2g' \Lm} + 1 + \Wm} , \\
\Pone 
&= e^{e^{-2g' \Lp}(\Wp + 1)} \, e^{e^{-2g' \Lm}(\Wm + 1)} \bigg( \, (1 + \Wp + \Wp^2)(1 + \Wm + \Wm^2) \, e^{-4g' \tau_0}  \left( e^{\Wp + 1} - e^{\Wm + 1} \right)^2 \\
&\quad + \Wp \Wm \bigg( -\left(e^{-2g' \Lp} + e^{-2g' \Lm} + 2\right) e^{2 + \Wp + \Wm} + e^{2(\Wp + 1)} e^{-2g' \Lm} + e^{2(\Wm + 1)} e^{-2g' \Lp} \bigg) \bigg), \\
\Ptwo 
&= \Wp \Wm \, e^{2 + \Wp + \Wm} \left[
e^{2(1 + \Wp) e^{- 2g' \Lp}}(1 + e^{-2g' \Lp}) +
e^{2(1 + \Wm)  e^{2g' \Lm}}(1 + e^{-2g' \Lm}) \right] \\
&\quad
- \Wp \Wm \, \left[e^{2(1 + \Wp + (1 + \Wm) e^{-2g' \Lm} - 2g' \Lm} +
e^{2(1 + \Wm + (1 + \Wp) e^{-2g \Lp} - 2g' \Lp}
\right] . 
\end{align*}
We plot the above expression for the transition probability in Fig. \ref{fig:9} from which one can see that the late time transition probability for the UDW detector along the Rindler-Rindler trajectory, and coupled to the massless real scalar field in the Minkowski vacuum state, is an increasing function of both $g$ and $g'$. Further, the transition probability asymptotically reaches the transition probability of a Rindler observer at acceleration $2g'$ in the Minkowski vacuum. The question of how fast the transition probability becomes asymptotically constant depends on both $g$ and $g'$. One can see this in the right panel, where the transition probability asymptotically approaches a constant value, the same as the Rindler one, at different times, which can be attributed to the choice of $g$ and $g'$. 
In the Rindler--Rindler case, the trajectory is nonstationary, so the correlators are not exactly proper-time translation invariant, and no global KMS condition is expected. The resulting thermality should therefore be understood only as an asymptotic or effective thermality at late times, when the trajectory approaches a uniformly accelerated trajectory with acceleration.

\subsection{Rindler vacuum} \label{tranprobrind}
We considered three types of trajectories: inertial, uniformly accelerated, and Rindler-Rindler, with a UDW detector coupled to a state that is vacuum with respect to the inertial observer. Now, we repeat the analysis for a state that is vacuum with respect to the Rindler observer, namely the Rindler vacuum. We begin with a UDW detector along an inertial trajectory $(t_0,x_0)$ = $(\tau,C)$, with $C > 1/g$ in the Rindler vacuum. The assumption $C > 1/g$ has been made so that the inertial trajectory lies within the right Rindler-Rindler wedge at $\tau = 0$ and leaves the wedge at $\tau_c \equiv \tau = C - 1/g$. For the comparison, all trajectories should lie in the same wedge. The pullback of the two-point function, for the Rindler vacuum, along the trajectory of an inertial observer, is given by
\begin{align}
 \mathcal{A} ^{\text{R}}_{\text{inertial}} (\tau, \tau') &= 
-\frac{1}{4 \pi} \left[
\frac{1}{
(C + \tau)
(C + \tau')
\left( \log(C + \tau) - 
       \log(C + \tau') -i\epsilon \right)^2}
\right. +  \left.
\frac{1}{
(C - \tau)
(C - \tau')
\left( \log(C - \tau) - 
       \log(C - \tau') -i\epsilon \right)^2}
\right].
\end{align}
The above two-point function $\mathcal{A} ^{\text{R}}_{\text{inertial}}$ is independent of $g$, and fully determined by $C$ and $\tau$, which determines the distance from the Rindler horizon. The dependence on \(C\), the spatial coordinate of the inertial detector, and on the inertial frame time \(\tau\) can be understood from the fact that the Rindler vacuum is not invariant under spacetime translations from the perspective of an inertial observer. In the limit $C \rightarrow \infty$, which correspond to being far away from the horizon, the above two point function $\mathcal{A} ^{\text{R}}_{\text{inertial}} \rightarrow \mathcal{A} ^{\text{M}}_{\text{inertial}} $. Therefore, far from the horizon, all properties that depend on the pullback of the two-point function along the detector's trajectory will be the same for both cases.  For definiteness, we compute the transition probability of the UDW detector moving along the inertial trajectory in the Rindler vacuum, which, under the approximation discussed in Appendix[\ref{Appendix C}] with vanishing residue contribution, is obtained to be
\begin{align*}
\mathcal{L}^{\text{R}}_{\text{inertial}} &= 
-\frac{\sigma^2 \lambda^2 e^{-\Omega^2 \sigma^2 / 2}}{4} \left[
\frac{1}{
(C + \tau_0 + \frac{i \Omega \sigma^2}{2})
(C + \tau_0 - \frac{i \Omega \sigma^2}{2})
\left( \log(C + \tau_0 + \frac{i \Omega \sigma^2}{2}) - 
       \log(C + \tau_0 - \frac{i \Omega \sigma^2}{2}) \right)^2}
\right. \\
& \qquad \left.
+ \frac{1}{
(C - \tau_0 - \frac{i \Omega \sigma^2}{2})
(C - \tau_0 + \frac{i \Omega \sigma^2}{2})
\left( \log(C - \tau_0 - \frac{i \Omega \sigma^2}{2}) - 
       \log(C - \tau_0 + \frac{i \Omega \sigma^2}{2}) \right)^2}
\right]. \numberthis \label{eq:RindlervacuumInertial}
\end{align*}
The above expression for $\mathcal{L}^{\text{R}}_{\text{inertial}}$ reduces to $\mathcal{L}^{\text{M}}_{\text{inertial}}$ in the limit $C \rightarrow \infty$, which corresponds to being far from the Rindler horizon, or to the late-time limit $\tau_0 \rightarrow \infty$. The transition probability for the Rindler trajectory in the Rindler vacuum, under the approximation discussed in Appendix[\ref{Appendix C}] with vanishing residue contribution, is 
\begin{align}
\mathcal{L}^{\text{R}}_{\text{Rindler}} &=
\frac{\lambda^2 e^{-\Omega^2 \sigma^2/2}}{2 \, \Omega^2 \sigma^2},
\end{align}
which is the same as the transition probability for an inertial UDW detector in Minkowski vacuum. Further, the transition probability of a UDW detector along Rindler-Rindler trajectory in the Rindler vacuum is given by
\begin{align}
\mathcal{L}^{\text{R}}_{\text{RRindler}} &=
\lambda^2 \sigma^2 e^{-\Omega^2 \sigma^2/2} 
\frac{\text{N}}
     {2 \, D}.
\end{align}
where,
\begin{align}
D &=
(e^{2g' y} + e^{2g' y_1} 
- 2 e^{g'(y + y_1)} \cos(g' \Omega \sigma^2))^2, \\
\text{N} &=
g'^2 e^{g'(y + y_1)} \bigg[
- (1 + \dot{y} \dot{y}_1) \cos(g' \Omega \sigma^2) 
(e^{2g' y} + e^{2g' y_1}) \\
&\qquad + 2(1 + \dot{y} \dot{y}_1) e^{g'(y + y_1)} 
+ (\dot{y} + \dot{y}_1) \sinh(i g' \Omega \sigma^2) 
(e^{2g' y} - e^{2g' y_1})
\bigg].
\end{align}
\begin{figure}
    \centering
    \includegraphics[width=.92\textwidth]{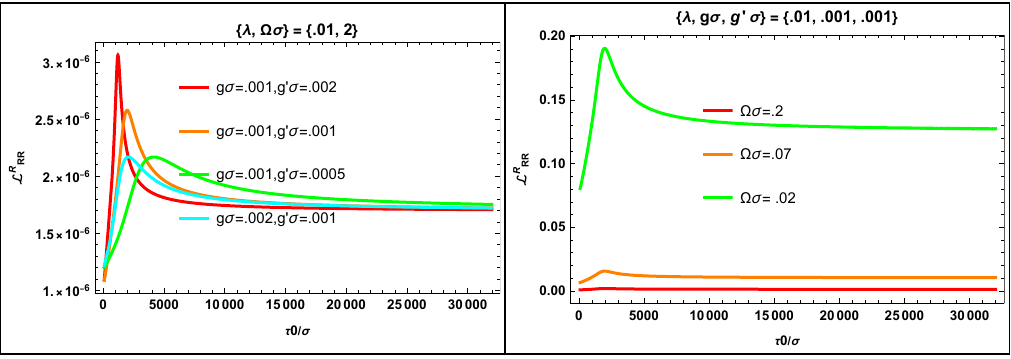}
     \captionsetup{margin=1cm, font=small}
    \caption{The above plots illustrate the transition probability of a UDW detector along the Rindler-Rindler trajectory interacting with a real massless scalar field in the Rindler vacuum in 1+1 spacetime dimensions, as discussed in Section \ref{tranprobrind}. } 
    \label{fig:10}
\end{figure}
Here, $y \equiv y(\tau_0+i \Omega\sigma^2/2)$ and $y_1 \equiv y(\tau_0-i \Omega\sigma^2/2)$. We plot the transition probabilities for the UDW detector in the Rindler vacuum in Fig.\ref{fig:10}. We see that for the Rindler-Rindler trajectory in Rindler vacuum, the transition probability asymptotically reaches that of a Rindler observer at acceleration $2 g'$ in Minkowski vacuum (refer to Eq.\eqref{eq:MinkvacRind} and Fig.\ref{fig:10}). This is expected since the Rindler-Rindler trajectory asymptotically reaches a shifted Rindler trajectory at late times.

\subsection{Rindler-Rindler vacuum} \label{tranprobRrind}
The conformal symmetry in (1+1) dimensions allows us to define a conformal vacuum for the nth Rindler transformation. This vacuum is non-stationary for the $n=2$ transformation, and is defined corresponding to the annihilation operator for the field modes having negative frequency with respect to the conformal time $t_2$. One can follow \cite{confvacuum, deSitter} for a discussion of the conformal vacuum.   
In this subsection, we investigate UDW detectors along inertial, uniformly accelerated, and Rindler-Rindler trajectories coupled to the Rindler-Rindler conformal vacuum. 

The trajectory of a Rindler observer is given by $(x_1,t_1)$ = $(c_1,\tau)$. In the Minkowski plane, this corresponds to 
\begin{align}
(x_0,t_0) = \bigg(\frac{e^{g c_1}}{g} \cosh{g \tau},\frac{e^{g c_1}}{g} \sinh{g \tau} \bigg).
\end{align}
Therefore, at $\tau = 0$, the Rindler observer is at $x_0 = e^{g c_1}/g$ and it will be inside the Rindler Rindler wedge if $e^{g c_1}/g > 2/g$ (i.e., $e^{g c_1} > 2$). However, in the Rindler plane ($t_1,x_1$) the Rindler observer will leave the Rindler-Rindler wedge when $\tau > c_1 $. So, we consider the Gaussian peak of the switching function to be much smaller than $c_1$.  The two-point function for this case will be the same as  $\mathcal{A} ^{\text{R}}_{\text{inertial}} (\tau, \tau')$, except $C$ is replaced by $c_1$, as the observer moves along a constant Rindler spatial coordinate. The transition probability for the UDW detector along the Rindler trajectory in the Rindler-Rindler vacuum, under the approximation discussed in Appendix[\ref{Appendix C}], is obtained to be
\begin{align*}
\mathcal{L}^{\text{RR}}_{\text{Rindler}} &= 
-\frac{1}{4} \lambda^2 \sigma^2 e^{-\frac{1}{2} \Omega^2 \sigma^2}
\left[
\frac{1}{
(c_1 + \tau_0 + \frac{\mathrm{i}}{2} \Omega \sigma^2)
(c_1 + \tau_0 - \frac{\mathrm{i}}{2} \Omega \sigma^2)
\left( 
\log(c_1 + \tau_0 + \frac{\mathrm{i}}{2} \Omega \sigma^2) - 
\log(c_1 + \tau_0 - \frac{\mathrm{i}}{2} \Omega \sigma^2)
\right)^2
} \right. \\
&\quad \left.
+ \frac{1}{
(c_1 - \tau_0 - \frac{\mathrm{i}}{2} \Omega \sigma^2)
(c_1 - \tau_0 + \frac{\mathrm{i}}{2} \Omega \sigma^2)
\left( 
\log(c_1 - \tau_0 - \frac{\mathrm{i}}{2} \Omega \sigma^2) - 
\log(c_1 - \tau_0 + \frac{\mathrm{i}}{2} \Omega \sigma^2)
\right)^2
}
\right], \numberthis \\
\end{align*}
which is same as $\mathcal{L}^{\text{R}}_{\text{inertial}}$.

\begin{figure} [H]
    \centering
    \includegraphics[width=.92\textwidth]{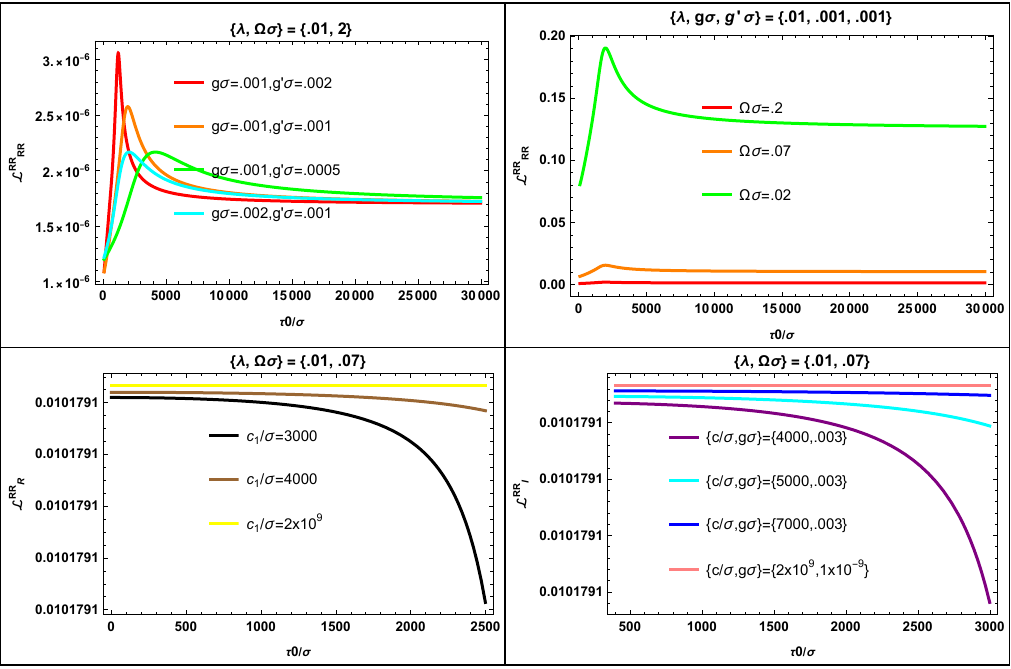}
     \captionsetup{margin=1cm, font=small}
    \caption{The above plots illustrate the transition probability of a UDW detector interacting with a real massless scalar field in the Rindler Rindler vacuum in 1+1 spacetime dimensions, as discussed in Section \ref{tranprobRrind}. } 
    \label{fig:11}
\end{figure}

Let us next consider a UDW detector in the Rindler-Rindler vacuum along an inertial trajectory $(t_0,x_0)$ = $(\tau,C)$, with $C > 1/g$. The assumption $C > 1/g$ has been made so that the inertial trajectory lies within the right Rindler-Rindler wedge at $\tau = 0$ and leaves the wedge at $\tau = C - 1/g$. The pullback of the momentum two-point function along the inertial trajectory in the Rindler-Rindler vacuum is obtained to be

\begin{align*}
\mathcal{A} ^{\text{RR}}_{\text{inertial}} (\tau, \tau') &= 
-\frac{1}{4 \pi} 
\left( \frac{1}{\left(C + \tau \right)
\left(C + \tau' \right)
\log\left( g \left(C + \tau \right) \right)
\log\left( g \left(C + \tau' \right) \right)  
\left[
\log\left(
\frac{\log\left[ g \left(C + \tau \right) \right]}{\log\left[ g \left(C +\tau' \right)   \right]}
\right) -i\epsilon \right]^2} + \right. \\
& \qquad \left.
 \frac{1}{\left(C - \tau \right)
\left(C - \tau' \right)
\log\left( g \left(C - \tau \right) \right)
\log\left( g \left(C - \tau' \right) \right)  
\left[
\log\left(
\frac{\log\left[ g \left(c - \tau \right) \right]}{\log\left[ g \left(c - \tau' \right) \right]}
\right) -i\epsilon \right]^2} \right). \numberthis
\end{align*}
The above two point function $\mathcal{A} ^{\text{RR}}_{\text{inertial}} $ is independent of $g'$ and fully determined by $g$, $C$, and $\tau$, which determines the distance from the bifurcation point.  The two point function $\mathcal{A} ^{\text{RR}}_{\text{inertial}} \rightarrow \mathcal{A} ^{\text{R}}_{\text{inertial}}$ in the limit $g \rightarrow 0$ \footnote{One can see it as $x \to 0^+$, we have $\ln(ax) \to -\infty$ and $\ln(bx) \to -\infty$ and
\[
\frac{\ln(ax)}{\ln(bx)} = \frac{1 + \frac{\ln a}{\ln x}}{1 + \frac{\ln b}{\ln x}} \to 1,
\]
so
\[
\ln\!\left(\frac{\ln(ax)}{\ln(bx)}\right) \approx \frac{\ln(a/b)}{\ln x}.
\]
Multiplying by $\ln(ax) \approx \ln x + \ln a$ gives
\[
\lim_{x \to 0^+} \ln(ax)\,\ln\!\left(\frac{\ln(ax)}{\ln(bx)}\right) = \ln\!\frac{a}{b}
\]
}. 
Furthermore, $\mathcal{A} ^{\text{RR}}_{\text{inertial}} \rightarrow \mathcal{A} ^{\text{M}}_{\text{inertial}}$ in the limit $C \rightarrow \infty$. Therefore, all properties that depend on the pullback of the two-point function along the trajectory of an observer will be identical in both cases in the appropriate limits. For definiteness, we compute the transition probability of the UDW detector moving along the inertial trajectory in the Rindler Rindler vacuum, which, under the approximation discussed in Appendix[\ref{Appendix C}] with vanishing residue contribution, is obtained to be

\begin{align}
\mathcal{L}^{\text{RR}}_{\text{inertial}} &= 
-\frac{1}{4} \lambda^2 \sigma^2 e^{-\frac{1}{2} \Omega^2 \sigma^2}
\left( \frac{1}{D_-}+ \frac{1}{D_+} \right),
\end{align}
where,
\begin{align*}
D_\pm &= 
\left(C \pm \tau_0 + \tfrac{\mathrm{i}}{2} \Omega \sigma^2\right)
\left(C \pm \tau_0 - \tfrac{\mathrm{i}}{2} \Omega \sigma^2\right)
\log\left( g \left(C \pm \tau_0 + \tfrac{\mathrm{i}}{2} \Omega \sigma^2 \right) \right)
\log\left( g \left(C \pm \tau_0 - \tfrac{\mathrm{i}}{2} \Omega \sigma^2 \right) \right) \times
\\
&\quad  
\left[
\log\left(
\log\left[ g \left(C \pm \tau_0 \pm \tfrac{\mathrm{i}}{2} \Omega \sigma^2 \right) \right]
\right)
-
\log\left(
\log\left[ g \left(C \pm \tau_0 \mp \tfrac{\mathrm{i}}{2} \Omega \sigma^2 \right) \right]
\right)
\right]^2 \numberthis
\end{align*}
Again the above expression for $\mathcal{L}^{\text{RR}}_{\text{inertial}}$ reduces to $\mathcal{L}^{\text{RR}}_{\text{Rindler}}$ = $\mathcal{L}^{\text{R}}_{\text{inertial}}$ in $g\rightarrow 0$ limit. Furthermore, in the limit $C \rightarrow \infty$ or $\tau_0 \rightarrow \infty$, $\mathcal{L}^{\text{RR}}_{\text{inertial}}$ reduces to $\mathcal{L}^{\text{M}}_{\text{inertial}}$ = $\mathcal{L}^{\text{R}}_{\text{Rindler}}$. One can also see these properties in the plots shown in Fig.\ref{fig:11}. The bottom left and bottom right panels of Fig. \ref{fig:11} are quite similar, since $g$ is small, though they correspond to the Rindler and inertial trajectories, respectively. Furthermore, the yellow curve of the bottom left panel and the pink curve of the bottom right panel of Fig.\ref{fig:11} look similar in magnitude to what one gets from Eqs.\eqref{eq:RindlervacuumInertial} and \eqref{eq:27a}, confirming that they match in the limit $C \rightarrow \infty$.

Denoting $y \equiv y(\tau_0+i \Omega\sigma^2/2)$ and $y_1 \equiv y(\tau_0-i \Omega\sigma^2/2)$, the transition probability of the UDW detector along the Rindler-Rindler trajectory in the Rindler-Rindler conformal vacuum, under the approximation discussed in Appendix[\ref{Appendix C}] with vanishing residue contribution, is given by 
\begin{align}
\mathcal{L}^{\text{RR}}_{\text{RRindler}} &= 
\frac{\lambda^2 \sigma^2 e^{-\frac{1}{2} \Omega^2 \sigma^2}}{2} 
\left[
\frac{
(1 + \dot{y} \dot{y}_1)\left( 
-(y - y_1)^2 - (\mathrm{i} \Omega \sigma^2)^2 \right) 
+ 2 (y - y_1) (\mathrm{i} \Omega \sigma^2)(\dot{y} + \dot{y}_1)
}{
\left( (y - y_1)^2 - (\mathrm{i} \Omega \sigma^2)^2 \right)^2
}
\right],
\end{align}
which is clearly time dependent and depends on both parameters $g$ and $g'$. From the upper panels of Fig.~\ref{fig:11}, it can be seen that the plots for $\mathcal{L}^{\text{RR}}_{\text{RRindler}}$ are similar to those for $\mathcal{L}^{\text{R}}_{\text{RRindler}}$ in Fig.~\ref{fig:10}, although their numerical values are slightly different. One can understand this from the fact that the Rindler-Rindler trajectory reduces to a shifted Rindler trajectory at late times.

\begin{figure} [H]
    \centering
    \includegraphics[width=.92\textwidth]{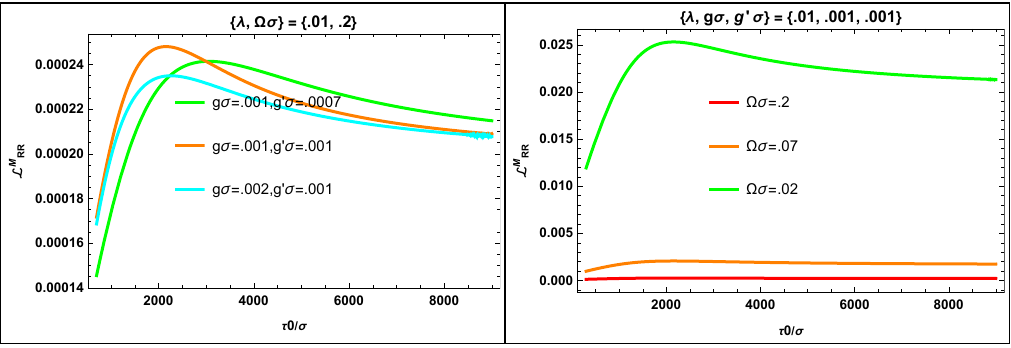}
     \captionsetup{margin=1cm, font=small}
    \caption{The above plots illustrate the transition probability of a UDW detector interacting with a real massless scalar field in the Minkowski vacuum in 3+1 spacetime dimensions, as discussed in Section \ref{sec:31trans}.} 
    \label{fig:minkvac}
\end{figure}

\section{Comparison with the 3+1 dimensional case} \label{sec:31trans}

Unlike the $(1+1)$-dimensional case discussed in the previous sections, the massless scalar field theory in $(3+1)$ dimensions is not conformally invariant. Consequently, the simplifications associated with conformal mode decompositions are no longer available. Nevertheless, for the present analysis of UDW detectors in Minkowski and Rindler vacua, it is not necessary to introduce a separate quantization associated with the Rindler--Rindler coordinates. The quantum fields prepared either in the standard Minkowski vacuum or in the standard Rindler vacuum are defined using stationary quantizations associated with their respective timelike Killing vectors.

For the Minkowski vacuum, the massless scalar field admits the standard plane-wave expansion
\begin{equation}
\hat{\phi}(x)=
\int \frac{d^3\mathbf{k}_0}
{(2\pi)^{3/2}\sqrt{2\omega_{\mathbf{k}_0}}}
\left(
\hat{a}_{\mathbf{k}_0}e^{-ik_0\cdot x}
+
\hat{a}^{\dagger}_{\mathbf{k}_0}e^{ik_0\cdot x}
\right),
\end{equation}
where the annihilation operators satisfy
\begin{equation}
\hat{a}_{\mathbf{k}_0} |0_M\rangle = 0.
\end{equation}
The corresponding Minkowski Wightman function is
\begin{equation}
W_M(x,x')
=
\langle 0_M |
\hat{\phi}(x)\hat{\phi}(x')
|0_M\rangle
=
-\frac{1}{4\pi^2}
\frac{1}
{
(t_0-t'_0-i\epsilon)^2
-
|\mathbf{x}_0-\mathbf{x}'_0|^2
}.
\label{WM4D}
\end{equation}

Similarly, one may define the Rindler vacuum by quantizing the field with respect to the boost Killing vector in the right Rindler wedge. The field operator may then be expanded as
\begin{equation}
\hat{\phi}(x)
=
\int_0^\infty d\omega_1
\int d^2k_{1\perp}
\left[
\hat{b}_{\omega_1 k_{1\perp}}
u_{\omega_1 k_{1\perp}}(x)
+
\hat{b}^{\dagger}_{\omega_1 k_{1\perp}}
u^*_{\omega_1 k_{1\perp}}(x)
\right],
\end{equation}
where the mode functions
$u_{\omega_1 k_{1\perp}}(x)$
have positive frequency with respect to the Rindler time coordinate. The Rindler vacuum $|0_R\rangle$ is defined through
\begin{equation}
\hat{b}_{\omega_1 k_{1\perp}}|0_R\rangle =0.
\end{equation}
The corresponding Wightman bidistribution is\cite{candelas1976quantum}
\begin{equation}
W_R(x,x')
=
\langle 0_R|
\hat{\phi}(x)\hat{\phi}(x')
|0_R\rangle = W_{M}(\mathbf{x},\mathbf{x}')
-
\int_{-\infty}^{\infty}
\frac{dv}{\pi^{2}+v^{2}}\,
W_{M}\!\bigl(\mathbf{x},\mathbf{x}''(v)\bigr).
\end{equation}
where $W_{M}(\mathbf{x},\mathbf{y})$is the Minkowski vacuum Wightman bidistribution, and the points $\mathbf{x}=(t_0,x_0,y_0,z_0),
\qquad
\mathbf{x}'=(t'_0,x'_0,y'_0,z'_0)$ are in the right-hand-side Rindler wedge, with $\mathbf{x}''(v)
:= \left( -\,t'_0\cosh v - x'_0\sinh v,\, -\,x'_0\cosh v - t'_0\sinh v,\, y'_0,\, z'_0 \right)$. 

Having specified the quantum state, one may evaluate detector observables along arbitrary trajectories $x^\mu(\tau)$. The nonstationarity relevant to the present discussion arises from the choice of trajectories rather than from the definition of the quantum state itself. In particular, the Rindler--Rindler trajectory is not generated by a timelike Killing flow, and therefore the pullback of the Wightman function along this trajectory, $W\big(x(\tau),x(\tau')\big),$
is generally not invariant under translations of the detector proper time. Consequently, the detector response along the Rindler--Rindler trajectory is intrinsically time dependent and need not satisfy an exact KMS condition at finite times, even when the quantum field is prepared in a stationary state such as the Minkowski or Rindler vacuum. The approach to a constant late-time transition rate should therefore not be interpreted as the existence of an exact thermal equilibrium associated with the full Rindler--Rindler trajectory. Rather, it reflects the fact that the trajectory asymptotically approaches a uniformly accelerated worldline with approximately constant proper acceleration.

\begin{figure}[H]
    \centering
    \includegraphics[width=.92\textwidth]{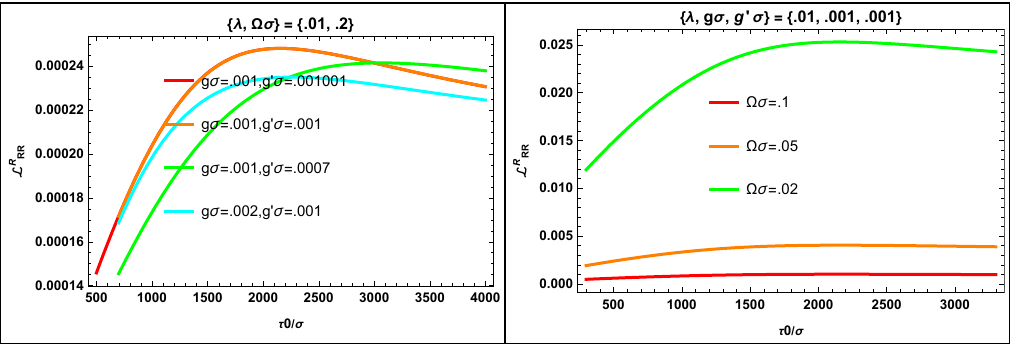}
     \captionsetup{margin=1cm, font=small}
    \caption{The above plots illustrate the transition probability of a UDW detector along the Rindler-Rindler trajectory interacting with a real massless scalar field in the Rindler vacuum in 3+1 spacetime dimensions, as discussed in Section \ref{sec:31trans}. } 
    \label{fig:rindvac3}
\end{figure}

To examine the effects of dimensionality and compare with the existing results in literature, we next consider a UDW detector coupled to the amplitude of a real, massless scalar field in 3+1-dimensional Minkowski spacetime, with the following interaction Hamiltonian: 
\begin{equation} \label{eq:22a}
    H_{\text{int}} = \lambda \chi(\tau) \hat{\mu} (\tau) \hat{\phi} (\textbf{x}(\tau)).
\end{equation}
Here, $\lambda$ is a small coupling constant, $\chi$ is the switching function, and $\hat{\mu}$ is the monopole moment operator of the detector. We take the same Gaussian switching function $\chi$ as in the previous section for computing the transition probability, similar to the previous section, and display the results in Fig.\ref{fig:minkvac} and Fig.\ref{fig:rindvac3}. Moreover, we show the comparison of the analytical saddle point approximation result with the numerical evaluation in Fig.\ref{fig:comparenumana}.

\begin{figure*}[ht!] 
        \centering
        \includegraphics[width=0.92\textwidth]{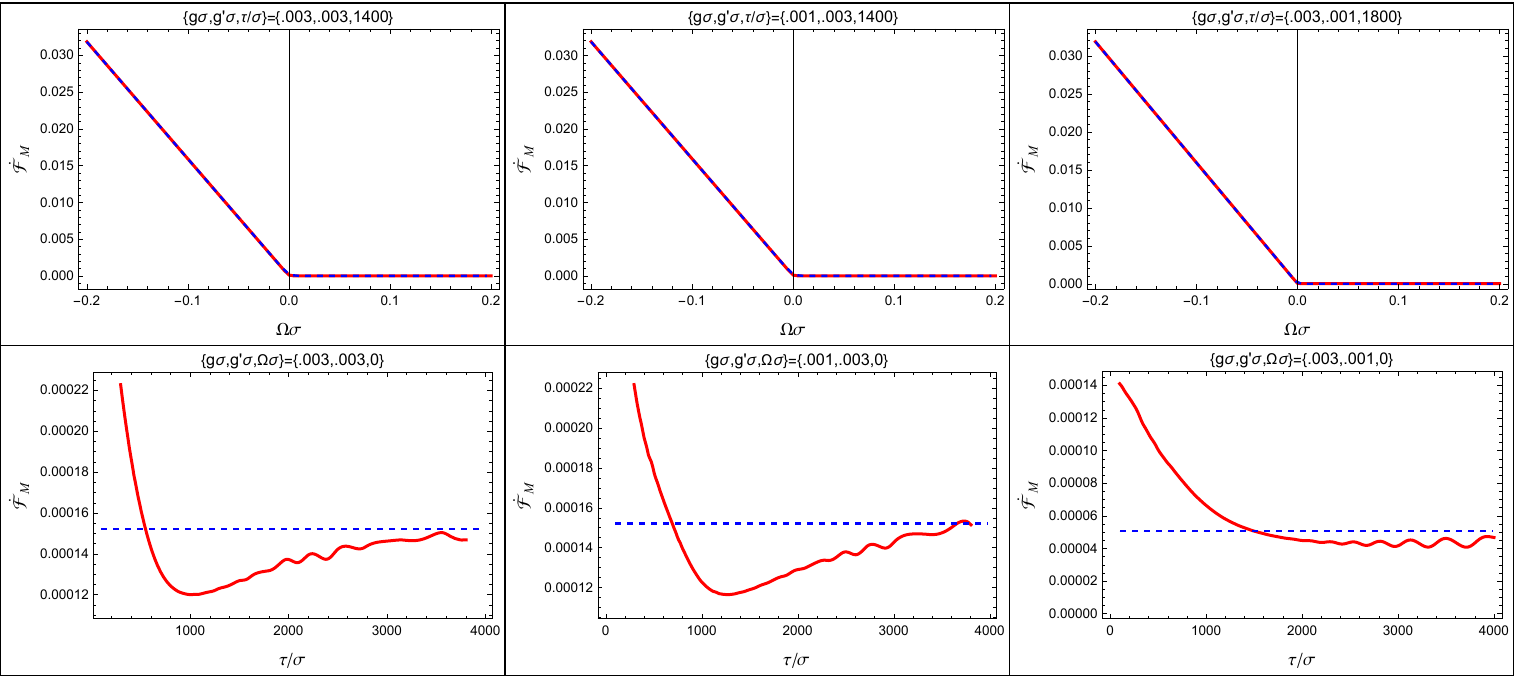}
        \captionsetup{margin=1cm, font=small}
        \caption{The above plots illustrate the transition rate of a UDW detector interacting with a real massless scalar field in the Minkowski vacuum in 3+1 spacetime dimensions, as discussed in Section \ref{sec:31trans}. The red curves represent the Rindler-Rindler trajectory, while the blue curves represent the Rindler trajectory, with acceleration $2g'$, far from the horizon.}    
        \label{fig:fourminkvaratecomball}
\end{figure*}

\begin{figure} 
    \centering
    \includegraphics[width=.92\textwidth]{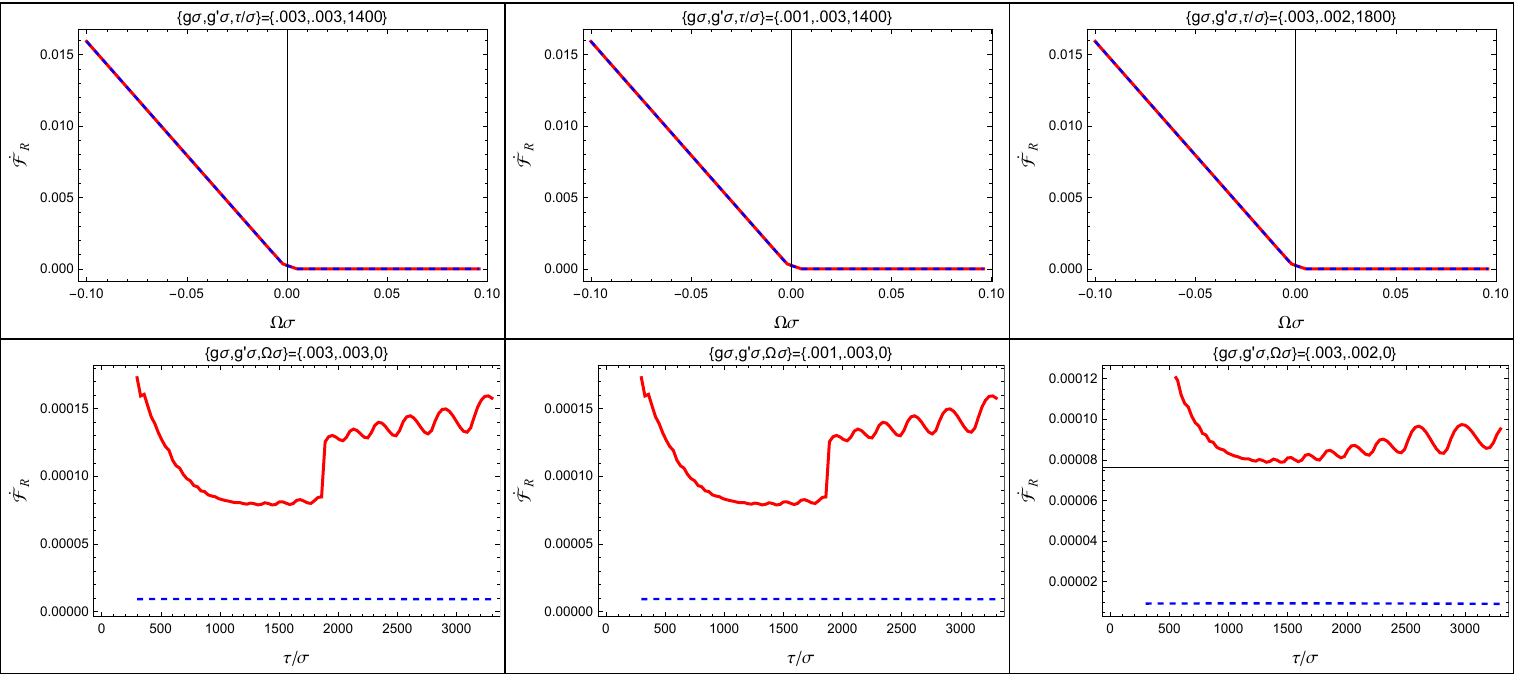}
     \captionsetup{margin=1cm, font=small}
    \caption{The above plots illustrate the transition rate of a UDW detector interacting with a real massless scalar field in the Rindler vacuum in 3+1 spacetime dimensions, as discussed in Section \ref{sec:31trans}. The red curves represent the Rindler Rindler trajectory, while the blue curves represent the inertial trajectory far from the horizon.} 
    \label{fig:rindvacrate}
\end{figure}

In addition to discussing the transition probability, we also compute the transition rate numerically, in parallel with the transition probability calculation. The expression for the transition rate is given in \cite{Louko:2006zv, Louko:2007mu}:
\begin{equation} \label{eq:23}
\dot{\mathcal{F}} (\Omega, \tau) = -\frac{\omega}{4\pi} 
+ 2 \int_{0}^{\Delta\tau} \! ds \, \Re \left[ 
e^{-i \omega s} W_{0}(\tau, \tau - s) 
+ \frac{1}{4 \pi^{2} s^{2}} \right] 
+ \frac{1}{2 \pi^{2} \Delta\tau} + O(\delta).
\end{equation}
Here, \(\delta\) denotes the small switching parameter that regulates the smoothness of the detector's coupling to the field, with the term \(O(\delta)\) representing corrections that vanish in the sharp-switching limit \(\delta \to 0\). The above expression, Eq.\eqref{eq:23}, refines the detector model by isolating finite, physically meaningful contributions from the field’s Wightman function while systematically removing short-distance divergences.  The inclusion of the counterterm \(1/(4\pi^{2}s^{2})\) within the integrand cancels the universal ultraviolet singularity of the Wightman function, yielding a well-behaved integrand even at coincident points. The term \(-\omega/(4\pi)\) serves as a vacuum subtraction ensuring consistency with the expected Minkowski-space response, while the finite-duration correction \(1/(2\pi^{2}\Delta\tau)\) accounts for the effects introduced by switching the detector on and off over a finite proper time. In the limit of long interaction time, Eq.\eqref{eq:23} converges to a steady-state transition rate consistent with thermal responses such as the Unruh and Hawking effects. We use Eq.\eqref{eq:23} for computing the transition rate and show the numerical results of the computation in Fig.\ref{fig:fourminkvaratecomball} and Fig.\ref{fig:rindvacrate}.

From the plots in Fig.\ref{fig:minkvac} and Fig.\ref{fig:fourminkvaratecomball}, one can observe that, for the Minkowski vacuum case, the transition rate and the transition probability for a detector following a Rindler--Rindler trajectory asymptotically approach those of a detector coupled to the Minkowski vacuum on a Rindler trajectory with acceleration \( 2g' \), provided that \( g < g' \) \footnote{The small oscillations observed around the mean value in the red curves shown in the bottom panels of Fig.~\ref{fig:fourminkvaratecomball} are likely attributable to the computational limitations associated with the Rindler–Rindler trajectory; however, they do not affect the overall physical behavior or the conclusions drawn from the analysis.
}. However, for the \( g > g' \) case, the rate asymptotically reaches a constant value whose magnitude depends upon both $g$ and $g'$. 
For both inertial and uniformly accelerated (Rindler) trajectories in the Minkowski vacuum, the transition probability of an amplitude-coupled Unruh--DeWitt detector in \(3+1\) dimensions equals \( 1/2\pi \) times the transition probability of a derivative-coupled Unruh--DeWitt detector in \(1+1\) dimensions along the respective trajectories, as described in equations~\eqref{eq:27a} and~\eqref{eq:MinkvacRind}. For the amplitude coupled UDW detector along the Rindler-Rindler trajectory in 3+1 dimensions, the numerical plots too suggest that the corresponding transition probability is also equal to \(1/2\pi \) times the transition probability of a derivative-coupled Unruh--DeWitt detector in \(1+1\) dimensions.

From the plots in Fig.\ref{fig:rindvac3} and Fig.\ref{fig:rindvacrate}, one can observe that, in the case of Rindler vacuum, the transition rate and transition probability for a detector following a Rindler--Rindler trajectory asymptotically approach those of a detector with acceleration \( 2g' \), provided that \( g < g' \). In contrast with the (1+1)-dimensional situation discussed above, the transition probability of the detector moving along the Rindler trajectory in the Rindler vacuum now depends on $g$. 
Furthermore, in contrast with the Minkowski vacuum, for the Rindler vacuum, the transition probabilities in 1+1 dimensions and 3+1 are not related by just a constant factor.
The transition rate of an inertial detector in the Rindler vacuum is given by (see \cite{Louko:2007mu})
\begin{equation}
     \dot{\mathcal{F}}_{\text{inertial}} ^\text{R} (\Omega, \tau) = \dot{\mathcal{F}}_{\text{inertial}} ^\text{M} (\Omega, \tau) +  \dot{\mathcal{F}}_{\text{inertial}} ^\text{extra} (\Omega, \tau)
\end{equation}
where,
\begin{equation}
    \dot{\mathcal{F}}_{\text{inertial}} ^\text{M} (\Omega, \tau) =  
-\frac{\omega}{4\pi} 
+ \frac{\cos(\omega \, \Delta \tau)}{2\pi^2 \, \Delta \tau} 
+ \frac{\omega}{2\pi^2} \, \text{Si}(\omega \, \Delta \tau),
\end{equation}
with Si representing the SinIntegral and 
\begin{equation}
    \dot{\mathcal{F}}_{\text{inertial}} ^\text{extra} (\Omega, \tau) =
\frac{1}{2\pi^2} \int_0^{\Delta \tau} 
\frac{\cos(\omega s)}{s^2} \left(
1 - \frac{s}{2\tau - s} 
\left( \frac{1}{\log\left(\frac{C - \tau}{C - \tau + s}\right)} 
+ \frac{1}{\log\left(\frac{C + \tau}{C + \tau - s}\right)} 
\right)
\right) ds.
\end{equation}
The second term in the above expression for $\dot{\mathcal{F}}_{\text{inertial}} ^\text{extra}$ is finite in the limit $s \rightarrow 0$, while logarithmically divergent in the limit $\tau \rightarrow C$. So, the transition rate now depends on time and the parameter $C$, which determines the distance from the horizon at a given time. 

\section{Summary and discussion} \label{sec:conclusion}

We investigated the restriction to a wedge in Rindler spacetime, for which the Rindler vacuum appears thermal, and extended this analysis to the conformal vacuum associated with a sequence of wedges obtained by applying $n$ successive Rindler-like transformations to the inertial co-ordinates in Minkowski spacetime. The Fock space calculation using Bogoliubov transformations reaffirms that the vacuum of the $(n-1)th$ Rindler observer appears thermal to the nth Rindler observer at the $t_0=0$ time slice.

To understand the relation between different wedges from the perspective of a local observer, we examined the characteristic trajectories confined to these wedges, computed their corresponding accelerations, and analyzed the resulting horizon shifts arising from such restrictions. For the case $n = 2$, referred to as the \emph{Rindler--Rindler trajectory}, we found that the late-time acceleration asymptotically approaches $2g'$ when the velocity in the Rindler--Rindler plane, $dy/d\tau$, is negative, whereas for positive velocity in this plane, the late-time acceleration diverges. Numerical analysis further indicates that, for the general $n$-th Rindler case, the late-time acceleration asymptotically approaches a constant value for an appropriate branch of the solution. Additionally, we observed that the horizon experienced by a Rindler--Rindler observer is displaced from the standard Rindler horizon by an amount $1/g$, while for the general $n$-th Rindler observer, this horizon shift depends on the set of parameters $\{ g_1, g_2, \ldots, g_{n-1} \}$.

We coupled a Unruh–DeWitt (UDW) detector to a massless real scalar field prepared in different states and evaluated the detector’s response along these trajectories. The transition probability, evaluated via the saddle-point approximation, for a UDW detector following the Rindler–Rindler trajectory in Minkowski vacuum is found to coincide with that of a detector on a standard Rindler trajectory with acceleration equal to twice the parameter of the second Rindler transformation (i.e, $2g'$). Moreover, the late-time transition rate, computed numerically, is found to be Planckian, as expected, since at late times the acceleration asymptotically reaches a constant value \footnote{It should be noted that, in general, the response of the detector depends on the form of the switching function. However, since the trajectory asymptotically approaches a uniformly accelerated trajectory, numerical computations suggest that, under appropriate conditions, smooth switching functions, peaked in the late-time regime, yields thermality.}. This is in contrast with the Bogoliubov calculation result for the expectation value of the number operator at $t_0=0$ in Eq.(28) of \cite{Kolekar:2013hra} that is not Planckian and depends on both parameters $g$ and $g'$. In the Rindler vacuum, the transition probability of the Rindler Rindler observer asymptotically reaches the transition probability of a Rindler observer with acceleration $2g'$. Interestingly, the state that appears as a vacuum to the Rindler Rindler observer is experienced by a Rindler observer much like the way an inertial observer perceives the Rindler vacuum when far from the horizon. Furthermore, an inertial observer in the Rindler Rindler vacuum with $g \to 0$ perceives the same physics as an inertial observer in the Rindler vacuum when far from the horizon.

In the case of the Minkowski vacuum, the transition probability for momentum-coupled UDW detectors in $1+1$ dimensions differs from the corresponding result with an amplitude-coupled detector in $3+1$ dimensions only by an overall factor of $4\pi$. We found it to hold good for all three trajectories, namely the inertial, Rindler and the Rindler trajectories. In contrast, in the Rindler vacuum, the transition probability doesn't have such a nice relation between $1+1$ and $3+1$ dimensions. However, even in the case of the Rindler vacuum in $3+1$ dimensions, the transition probability for a UDW detector following a Rindler-Rindler trajectory asymptotically approaches that of the detector with a uniform proper acceleration $2g'$.

The \emph{Rindler--Rindler framework} developed here provides a mathematically precise description of spacetimes exhibiting nested acceleration scales and opens several promising directions for future work. In analogue gravity models characterized by two distinct flow gradients, the background velocity field may support both a primary horizon with surface gravity $\kappa_1$ and a secondary modulation with amplitude $\kappa_2$, varying as $\delta v(x) \sim e^{\alpha (x - x_h)}$ near the main horizon. Expanding the metric in tortoise-type coordinates about the primary horizon naturally yields a Rindler form governed by $\kappa_1$, while a subsequent Rindler-type transformation with parameter $g_2 \sim \alpha$ maps the system into a \emph{Rindler--Rindler metric}---a conformally Rindler spacetime encoding hierarchical acceleration structure. The conformal flatness of this construction ensures that field equations remain analytically tractable, enabling explicit determination of mode functions and detector response.  

Beyond its theoretical consistency, the hierarchical framework of $n^{th}$ Rindler provides a natural setting to explore multi-scale thermality and vacuum perception in curved or effectively curved spacetimes. For instance, the Rindler--Rindler trajectory can be interpreted as modeling a dynamical process where a black hole initially present at $v=0$ completely evaporates by $v=\infty$, allowing for an explicit mapping between nested acceleration horizons and evolving causal structures. This may also offer new insights into time-dependent Hawking-like processes and the correspondence between mirror radiation, electron emission, and black hole evaporation, extending the ideas of \cite{PhysRevD.94.065010, Lin:2024ihr}. Furthermore, investigating higher-order ($n>2$) Rindler hierarchies could reveal a rich hierarchy of ``thermalizations within thermalizations,'' potentially illuminating how effective temperatures emerge in non-inertial quantum systems with multiple acceleration scales. Finally, incorporating backreaction effects, interactions, or curved background geometries may extend the applicability of the Rindler--Rindler formalism to semiclassical gravity, quantum information flow across multiple horizons, and emergent spacetime scenarios in analogue and cosmological contexts.

\begin{appendices}

\section{Shift of $n^{th}$ Rindler} \label{appenA}

\begin{align*}
x_0 - t_0 &= \frac{1}{g_1}  
    \exp\left(\frac{g_1}{g_2} e^{g_2 x_2} \cosh(g_2 t_2)\right)  \exp\left(-\frac{g_1}{g_2} e^{g_2 x_2} \sinh(g_2 t_2)\right) 
\\[6pt]
&= \frac{1}{g_1}  
    \exp\left(\frac{g_1}{g_2} e^{g_2 x_2}(\cosh(g_2 t_2) - \sinh(g_2 t_2))\right) 
 \\[6pt]
&= \frac{1}{g_1} 
    \exp\left(\frac{g_1}{g_2} \exp\big(g_2(x_2 - t_2)\big)\right) 
 \\[6pt]
&= \frac{1}{g_1} 
    \exp\left(\frac{g_1}{g_2} \exp\Big(\frac{g_2}{g_3} e^{g_3 x_3}(\cosh(g_3 t_3) - \sinh(g_3 t_3))\Big)\right) 
 \\[6pt]
&= \frac{1}{g_1}  
    \exp\left(\frac{g_1}{g_2} \exp\Big(\frac{g_2}{g_3} e^{g_3(x_3 - t_3)}\Big)\right) 
\\[6pt]
&= \frac{1}{g_1}  
    \exp\left(\frac{g_1}{g_2} \exp\Big(\frac{g_2}{g_3} 
    \exp\big(\tfrac{g_3}{g_4} \exp(\tfrac{g_4}{g_5} \exp(\cdots \exp(\tfrac{g_{n-1}}{g_n} e^{g_n(x_n - t_n)})))\big)\Big)\right) 
\end{align*}
Substituting $x_n \approx t_n$ one gets
\begin{equation} \label{eqs:shift}
   x_0 -t_0 = \frac{1}{g_1}  
    \exp\left(\frac{g_1}{g_2} \exp\Big(\frac{g_2}{g_3} 
    \exp\big(\tfrac{g_3}{g_4} \exp(\tfrac{g_4}{g_5} \exp(\cdots \exp(\tfrac{g_{n-1}}{g_n})))\big)\Big)\right) ,
\end{equation}
which gives the shift for the $n^{th}$ Rindler. In particular, for Rindler the above expression gives $x_0-t_0 = 1/g$, which is the point where the Rindler trajectory cuts the $x_0$-axis. However, for large times when $x_n<<t_n$ and $t_n \rightarrow \infty$,
\begin{equation} \label{horizonshift}
  x_0 -t_0 \approx  \frac{1}{g_1}  
    \exp\left(\frac{g_1}{g_2} \exp\Big(\frac{g_2}{g_3} 
    \exp\big(\tfrac{g_3}{g_4} \exp(\tfrac{g_4}{g_5} \exp(\cdots \exp(\tfrac{g_{n-2}}{g_{n-1}})))\big)\Big)\right) ,
\end{equation}
which is 0 for the Rindler (since for Rindler $x_1 =0$), and $1/g_1$ for the Rindler-Rindler. 

\section{ Acceleration }  \label{Appendix B}
The time component of proper accleration along the Rindler Rindler trajectory is given by
\begin{align}
a^{0} &= \Gamma^{0}_{00} \left( \frac{dt_2}{d\tau} \right)^{2}
      + 2 \Gamma^{0}_{01} \frac{dt_2}{d\tau} \frac{dx_2}{d\tau}
      + \Gamma^{0}_{11} \left( \frac{dx_2}{d\tau} \right)^{2} \\
     & = g g' t_1 + 2 (gg'x_1 +g')\dot{y} + g g' t_1\dot{y}^{2}.
\end{align}
One can obtain the following spacelike component of the proper acceleration by differentiating \( u^{\mu} u_{\mu} = -1 \), 
\begin{align}
a^{1} = \frac{ gg' t_1}{\dot{y}} + 2 (gg'x_1+g')+ gg't_1 \dot{y}
\end{align}
Substituting $a^{0}$ and $a^{1}$ components computed above in $a = \sqrt{a^{\mu}a_\mu}$, we get the following expression for the proper acceleration
\begin{align}
a &= g' \left( \frac{gt_1 (1 + \dot{y}^2)}{\dot{y}} + 2 (g \, x_1 + 1) \right) \\[6pt]
&= 2 g'+  \frac{g }{2 \dot{y}} \left( e^{g'(y+ \tau)} (\dot{y}+1)^2 - e^{g'(y - \tau)} (\dot{y}-1)^2 \right)
\end{align}
For the negative root of $\dot{y}$, at late times ($\tau \rightarrow \infty$), the first two terms dominate, so
\begin{align}
a \approx 2 g'+  \frac{g }{2 \dot{y}} \left( e^{g'(y+ \tau)} (\dot{y}+1)^2 \right)
\end{align}
Substituting the late time trajectory obtained in Eq.\eqref{eq:15} the proper acceleration becomes
\begin{equation} 
    a(\tau) = 2g' - \frac{2g'}{W\!\left( \frac{g^2 e^{2g'\tau-1}}{4g'^2} \right)} - \ldots ,
\end{equation}
For the positive root of $\dot{y}$, the corresponding late time trajectory is given by Eq.~\eqref{eq:21}. Following the same procedure as above, we obtain the following expression for the proper acceleration:
\begin{eqnarray}
    a & = 2 g' + \frac{g}{2(1 + dS/d\tau)} e^{g'(2\tau + S) } (2 + dS/d\tau)^2 
\end{eqnarray}
where,
\begin{eqnarray}
    S = \sum _{k=1} ^\infty \frac{(-1/4 g')^k}{k !} \bigg(Ei \bigg( - \frac{g}{g'} e^{g' (y + \tau)} \bigg) ^k \bigg)^{(k-1)} \bigg|_{y=\tau}.
\end{eqnarray}
Taking late times limit one gets $\lim_{\tau \rightarrow \infty} a = \infty$.

\section{Computing transition probability} \label{Appendix C}
In this section, we introduce the saddle point approximation method for evaluating the transition probability, expressed as a double integral involving the switching functions and the two-point function $\mathcal{A}$ \cite{Salton2015, ARFKEN2013551, PhysRevD.102.085013}, given by  
\begin{align}
     L  &= \lambda^2 \int_{-\infty}^{+\infty} d\tau \int_{-\infty}^{+\infty} d\tau' \, 
     \chi(\tau)\chi(\tau') e^{- i \Omega (\tau - \tau')} \mathcal{A}(x(\tau),x'(\tau')).
\end{align}
Choosing a Gaussian profile for the switching function centered at $\tau_0$, the above expression takes the form  
\begin{align}
      L &= \lambda^2 \int_{-\infty}^{+\infty} d\tau \int_{-\infty}^{+\infty} d\tau' 
      \, e^{- \frac{(\tau - \tau_0)^2}{\sigma^2}} e^{- \frac{(\tau' - \tau_0)^2}{\sigma^2}} 
      e^{- i \Omega (\tau - \tau')} \mathcal{A}(x(\tau),x'(\tau')) \label{eq:requiredmiddlestep} \\
       &= \lambda^2 e^{-\frac{\sigma^2 \Omega^2}{2}} \int_{-\infty}^{+\infty} d\tau \int_{-\infty}^{+\infty} d\tau' 
      \, e^{- \frac{(\tau - \tau_0 + i \Omega \sigma^2 /2)^2}{\sigma^2}} 
      e^{- \frac{(\tau' - \tau_0 - i \Omega \sigma^2 /2)^2}{\sigma^2}} 
      \mathcal{A}(x(\tau),x'(\tau')).
\end{align}
Introducing the symmetric and antisymmetric combinations $\tilde{x} = \tau + \tau'$ and $\tilde{y} = \tau - \tau'$, the integral reduces to  
\begin{eqnarray*}
    L = \lambda^2 e^{-\frac{\sigma^2 \Omega^2}{2}} \int_{-\infty}^{+\infty} d\tilde{x} \int_{-\infty}^{+\infty} d\tilde{y} \,
    e^{- \frac{(\tilde{y} + i \Omega\sigma^2)^2}{2\sigma^2}} 
    e^{- \frac{(\tilde{x} - 2\tau_0)^2}{2\sigma^2}} 
    \mathcal{A}\!\left(x\!\left(\tfrac{\tilde{x}+\tilde{y}}{2}\right), x'\!\left(\tfrac{\tilde{x}-\tilde{y}}{2}\right)\right).
\end{eqnarray*}
Shifting the contour for $\tilde{y}$ by $\Omega\sigma^2$ eliminates the imaginary part of the Gaussian factors. The dominant contribution is then extracted via the saddle–point method, leading to  
\begin{equation} \label{eq:66}
    L \approx \pi \sigma^2 \lambda^2 e^{-\sigma^2 \Omega^2/2} 
    \mathcal{A}(x(\tau_0 + i \Omega \sigma^2/2), x'(\tau_0 - i \Omega \sigma^2/2)) 
    + \text{residual terms}.
\end{equation}
Under the assumption that no poles are crossed during the contour deformation the residual contributions vanish.

\subsection{Summing the residue for the Rindler observer} \label{App:sumResidue}
We can sum the residues and evaluate the limit $\sigma \rightarrow \infty$ for a few special cases. Substituting the two point function for the Rindler observer in Minkowski vacuum, Eq.\eqref{eq:Rindlermomntcorr}, into the detector response Eq.\eqref{eq:requiredmiddlestep} and introducing the variables
\[
u=\tau-\tau',
\qquad
s=\frac{\tau+\tau'}{2},
\]
with Jacobian \(d\tau\,d\tau'=ds\,du\), and
\[
(\tau-\tau_0)^2+(\tau'-\tau_0)^2
=
2(s-\tau_0)^2+\frac{u^2}{2}.
\]
we get the following response function\footnote{We have used the fact that the integral over \(s\) is Gaussian:
\[
\int_{-\infty}^{\infty} ds\,
e^{-2(s-\tau_0)^2/\sigma^2}
=
\sqrt{\frac{\pi}{2}}\,\sigma.
\]}

\begin{equation}
\LmRindler
=
-\frac{\lambda^2 g^2}{8\pi}
\sqrt{\frac{\pi}{2}}\sigma
\int_{-\infty}^{\infty}du\,
\frac{
e^{-u^2/(2\sigma^2)}e^{-i\Omega u}
}{
\sinh^2\!\left[\frac g2(u-i\epsilon)\right]
}.
\label{eq:start}
\end{equation}

Completing the square in the exponent,
\begin{align}
-\frac{u^2}{2\sigma^2}-i\Omega u
&=
-\frac{1}{2\sigma^2}
\left(u+i\Omega\sigma^2\right)^2
-\frac12\Omega^2\sigma^2 .
\end{align}

Hence,
\begin{equation}
\LmRindler
=
-\frac{\lambda^2 g^2}{8\pi}
\sqrt{\frac{\pi}{2}}\sigma\,
e^{-\Omega^2\sigma^2/2}
\int_{-\infty}^{\infty}du\,
\frac{
e^{-(u+i\Omega\sigma^2)^2/(2\sigma^2)}
}{
\sinh^2\!\left[\frac g2(u-i\epsilon)\right]
}.
\label{eq:shifted}
\end{equation}

The saddle point is located at
\begin{equation}
u_s=-i\Omega\sigma^2 .
\end{equation}

To evaluate the integral by steepest descent, deform the contour from
$\operatorname{Im}(u)=0$
to
$\operatorname{Im}(u)=-\Omega\sigma^2$.
During this deformation, the contour crosses the poles of
$\sinh^{-2}\!\left[\frac g2(u-i\epsilon)\right]$
located at
\begin{equation}
u_n=i\epsilon-\frac{2\pi i n}{g},
\qquad n=1,2,3,\dots
\end{equation}
which are second-order poles. The contour crosses all poles satisfying
\begin{equation}
\frac{2\pi n}{g}<\Omega\sigma^2+\epsilon.
\end{equation}
Therefore, we can write the total response as a sum of the saddle contribution and the contribution from poles:
\begin{equation}
\LmRindler
=
\mathcal L_{\text{saddle}}
+
\mathcal L_{\text{poles}} .
\end{equation}

\paragraph{Saddle contribution.}
Shift the integration variable to the saddle,
\begin{equation}
u=v-u_s=v-i\Omega\sigma^2 ,
\end{equation}
so that $(u+i\Omega\sigma^2)^2=v^2$ and $du=dv$. Then
\begin{equation}
\mathcal L_{\text{saddle}}
=
-\frac{\lambda^2 g^2}{8\pi}
\sqrt{\frac{\pi}{2}}\sigma\,
e^{-\Omega^2\sigma^2/2}
\int_{-\infty}^{\infty}dv\,
\frac{
e^{-v^2/(2\sigma^2)}
}{
\sinh^2\!\left[\frac g2\!\left(v-i\Omega\sigma^2-i\epsilon\right)\right]
}.
\end{equation}
The Gaussian localizes $v$ to $|v|\lesssim \sigma$, so to leading order one may
approximate the slowly varying denominator by its value at $v=0$,
\begin{equation}
\sinh^2\!\left[\frac g2\!\left(v-i\Omega\sigma^2-i\epsilon\right)\right]
\simeq
\sinh^2\!\left[-\frac{ig}{2}\!\left(\Omega\sigma^2+\epsilon\right)\right]
=
-\sin^2\!\left[\frac g2\!\left(\Omega\sigma^2+\epsilon\right)\right],
\end{equation}
giving
\begin{align}
\mathcal L_{\text{saddle}}
&\simeq
-\frac{\lambda^2 g^2}{8\pi}
\sqrt{\frac{\pi}{2}}\sigma\,
e^{-\Omega^2\sigma^2/2}
\left(
\frac{1}{-\sin^2\!\left[\frac g2(\Omega\sigma^2+\epsilon)\right]}
\right)
\int_{-\infty}^{\infty}dv\,e^{-v^2/(2\sigma^2)}
\nonumber\\
&=
\frac{\lambda^2 g^2}{8\pi}
\sqrt{\frac{\pi}{2}}\sigma\,
e^{-\Omega^2\sigma^2/2}
\frac{1}{\sin^2\!\left[\frac g2(\Omega\sigma^2+\epsilon)\right]}
\left(\sqrt{2\pi}\sigma\right)
\nonumber\\
&=
\frac{\lambda^2 g^2}{8}
\frac{\sigma^2}{\sin^2\!\left[\frac g2(\Omega\sigma^2+\epsilon)\right]}
e^{-\Omega^2\sigma^2/2}.
\end{align}
In particular,
\begin{equation}
\mathcal L_{\text{saddle}}
\sim
e^{-\Omega^2\sigma^2/2}
\to 0
\qquad (\sigma\to\infty).
\end{equation}

\paragraph{Pole contribution}
Define
\begin{equation}
f(u)=
e^{-u^2/(2\sigma^2)}e^{-i\Omega u}.
\end{equation}
Since \(\sinh\!\left[\tfrac g2(u-i\epsilon)\right]\) vanishes at poles, \(u=u_n\), we expand it to first order about \(u_n\):
\begin{equation}
\sinh\!\left[\frac g2(u-i\epsilon)\right]
=
\sinh\!\left[\frac g2(u-u_n)\right]
\simeq
\frac g2(u-u_n)\cosh\!\left(\frac g2(u_n-i\epsilon)\right)
=
(-1)^n\frac g2(u-u_n),
\end{equation}
so that
\begin{equation}
\frac{1}{\sinh^2\!\left[\frac g2(u-i\epsilon)\right]}
\simeq
\frac{4}{g^2}\frac{1}{(u-u_n)^2}.
\end{equation}
Since the poles are second order,
\begin{equation}
\operatorname{Res}_{u=u_n}
\frac{f(u)}{\sinh^2\!\left[\frac g2(u-i\epsilon)\right]}
=
\frac{4}{g^2}f'(u_n).
\end{equation}
Now,
\begin{align}
f'(u)
&=
\left(
-\frac{u}{\sigma^2}-i\Omega
\right)
f(u).
\end{align}
Substituting
\begin{equation}
u_n=i\epsilon-\frac{2\pi i n}{g},
\end{equation}
gives
\begin{align}
f'(u_n)
&=
-i\left(
\Omega-\frac{2\pi n}{g\sigma^2}
\right)
\exp\left[
\frac{(\epsilon-2\pi n/g)^2}{2\sigma^2}
+\Omega\epsilon
-\frac{2\pi n\Omega}{g}
\right]
\qquad .
\end{align}

Using the residue theorem for the downward contour shift (so that the crossed poles contribute with a minus sign),
\begin{align}
\mathcal L_{\text{poles}}
&=
-2\pi i
\left(
-\frac{\lambda^2 g^2}{8\pi}
\sqrt{\frac{\pi}{2}}\sigma
\right)
\sum_{n=1}^{N}
\frac{4}{g^2}f'(u_n)
\nonumber\\
&=
\sqrt{\frac{\pi}{2}}
\sigma\lambda^2
\sum_{n=1}^{N}
\left(
\Omega-\frac{2\pi n}{g\sigma^2}
\right)
\exp\left[
\frac{(\epsilon-2\pi n/g)^2}{2\sigma^2}
+\Omega\epsilon
-\frac{2\pi n\Omega}{g}
\right].
\end{align}
Here, $N$ is the number of crossed poles. Thus the full transition probability is
\begin{align}
\LmRindler
&\simeq
\frac{\lambda^2 g^2}{8}
\frac{\sigma^2}{\sin^2\!\left[\frac g2(\Omega\sigma^2+\epsilon)\right]}
e^{-\Omega^2\sigma^2/2}
\nonumber\\
&\quad+
\sqrt{\frac{\pi}{2}}
\sigma\lambda^2
\sum_{n=1}^{N}
\left(
\Omega-\frac{2\pi n}{g\sigma^2}
\right)
\exp\left[
\frac{(\epsilon-2\pi n/g)^2}{2\sigma^2}
+\Omega\epsilon
-\frac{2\pi n\Omega}{g}
\right].
\label{eq:full}
\end{align}

In the limit $\sigma\to\infty$,
\begin{equation}
\mathcal L_{\text{saddle}}
\sim
e^{-\Omega^2\sigma^2/2}
\to 0.
\end{equation}
For fixed $n$,
\begin{equation}
\frac{(\epsilon-2\pi n/g)^2}{2\sigma^2}\to 0,
\qquad
\frac{2\pi n}{g\sigma^2}\to 0,
\qquad
N\to\infty,
\end{equation}
and the pole sum is dominated by $n\ll \sigma^2$, so that the factor
$\exp\!\left(\frac{(\epsilon-2\pi n/g)^2}{2\sigma^2}\right)\to 1$ in the
dominant range. Therefore,
\begin{align}
\LmRindler
&\sim
\sqrt{\frac{\pi}{2}}
\sigma\lambda^2\Omega
\sum_{n=1}^{\infty}
e^{-2\pi n\Omega/g}.
\end{align}

Using
\begin{equation}
\sum_{n=1}^{\infty}e^{-an}
=
\frac{1}{e^a-1},
\end{equation}
with
\begin{equation}
a=\frac{2\pi\Omega}{g},
\end{equation}
we obtain
\begin{equation}
\LmRindler(\sigma\to\infty)
\sim
\sqrt{\frac{\pi}{2}}
\,\sigma\lambda^2
\frac{\Omega}{
e^{2\pi\Omega/g}-1
}.
\end{equation}

Hence the total transition probability diverges linearly with
$\sigma$,
while the transition rate remains finite and Planckian:
\begin{equation}
\Gamma
=
\lim_{\sigma\to\infty}
\frac{\LmRindler}{\sqrt{2\pi}\sigma}
=
\frac{\lambda^2}{2}
\frac{\Omega}{
e^{2\pi\Omega/g}-1
}.
\end{equation}

\end{appendices}

\bibliography{rref}

@article{Martinetti:2002sz,
    author = "Martinetti, Pierre and Rovelli, Carlo",
    title = "{Diamonds's temperature: Unruh effect for bounded trajectories and thermal time hypothesis}",
    eprint = "gr-qc/0212074",
    archivePrefix = "arXiv",
    doi = "10.1088/0264-9381/20/22/015",
    journal = "Class. Quant. Grav.",
    volume = "20",
    pages = "4919--4932",
    year = "2003"
}

@article{Kolekar:2013hra,
    author = "Kolekar, Sanved and Padmanabhan, T.",
    title = "{Quantum field theory in the Rindler-Rindler spacetime}",
    eprint = "1309.4424",
    archivePrefix = "arXiv",
    primaryClass = "gr-qc",
    doi = "10.1103/PhysRevD.89.064055",
    journal = "Phys. Rev. D",
    volume = "89",
    number = "6",
    pages = "064055",
    year = "2014"
}

@article{Padmanabhan:2019art,
    author = "Padmanabhan, T.",
    title = "{Gravity and Quantum Theory: Domains of Conflict and Contact}",
    eprint = "1909.02015",
    archivePrefix = "arXiv",
    primaryClass = "gr-qc",
    doi = "10.1142/S0218271820300013",
    journal = "Int. J. Mod. Phys. D",
    volume = "29",
    number = "01",
    pages = "2030001",
    year = "2019"
}

@article{confvacuum,
  title = {Conformal vacuum and the fluctuation-dissipation theorem in a de Sitter universe and black hole spacetimes},
  author = {Das, Ashmita and Dalui, Surojit and Chowdhury, Chandramouli and Majhi, Bibhas Ranjan},
  journal = {Phys. Rev. D},
  volume = {100},
  issue = {8},
  pages = {085002},
  numpages = {20},
  year = {2019},
  month = {Oct},
  publisher = {American Physical Society},
  doi = {10.1103/PhysRevD.100.085002},
  url = {https://link.aps.org/doi/10.1103/PhysRevD.100.085002}
}

@article{deSitter,
  title = {Quantum field theory in de Sitter and quasi--de Sitter spacetimes revisited},
  author = {Singh, Suprit and Ganguly, Chandrima and Padmanabhan, T.},
  journal = {Phys. Rev. D},
  volume = {87},
  issue = {10},
  pages = {104004},
  numpages = {15},
  year = {2013},
  month = {May},
  publisher = {American Physical Society},
  doi = {10.1103/PhysRevD.87.104004},
  url = {https://link.aps.org/doi/10.1103/PhysRevD.87.104004}
}

@article{Juarez-Aubry:2014jba,
    author = "Ju\'arez-Aubry, Benito A. and Louko, Jorma",
    title = "{Onset and decay of the 1 + 1 Hawking-Unruh effect: what the derivative-coupling detector saw}",
    eprint = "1406.2574",
    archivePrefix = "arXiv",
    primaryClass = "gr-qc",
    doi = "10.1088/0264-9381/31/24/245007",
    journal = "Class. Quant. Grav.",
    volume = "31",
    number = "24",
    pages = "245007",
    year = "2014"
}

@article{PhysRevD.104.065001,
  title = {General features of the thermalization of particle detectors and the Unruh effect},
  author = {Perche, T. Rick},
  journal = {Phys. Rev. D},
  volume = {104},
  issue = {6},
  pages = {065001},
  numpages = {14},
  year = {2021},
  month = {Sep},
  publisher = {American Physical Society},
  doi = {10.1103/PhysRevD.104.065001},
  url = {https://link.aps.org/doi/10.1103/PhysRevD.104.065001}
}

@article{PhysRev.115.1342,
  title = {Theory of Many-Particle Systems. I},
  author = {Martin, Paul C. and Schwinger, Julian},
  journal = {Phys. Rev.},
  volume = {115},
  issue = {6},
  pages = {1342--1373},
  numpages = {0},
  year = {1959},
  month = {Sep},
  publisher = {American Physical Society},
  doi = {10.1103/PhysRev.115.1342},
  url = {https://link.aps.org/doi/10.1103/PhysRev.115.1342}
}

@article{Louko:2006zv,
    author = "Louko, Jorma and Satz, Alejandro",
    title = "{How often does the Unruh-DeWitt detector click? Regularisation by a spatial profile}",
    eprint = "gr-qc/0606067",
    archivePrefix = "arXiv",
    doi = "10.1088/0264-9381/23/22/015",
    journal = "Class. Quant. Grav.",
    volume = "23",
    pages = "6321--6344",
    year = "2006"
}

@article{Louko:2007mu,
    author = "Louko, Jorma and Satz, Alejandro",
    title = "{Transition rate of the Unruh-DeWitt detector in curved spacetime}",
    eprint = "0710.5671",
    archivePrefix = "arXiv",
    primaryClass = "gr-qc",
    doi = "10.1088/0264-9381/25/5/055012",
    journal = "Class. Quant. Grav.",
    volume = "25",
    pages = "055012",
    year = "2008"
}

@article{Lochan:2021pio,
    author = "Lochan, Kinjalk and Padmanabhan, T.",
    title = "{A nested sequence of inequivalent Rindler vacua : Universal Relic Thermality of Planckian origin}",
    eprint = "2107.03406",
    archivePrefix = "arXiv",
    primaryClass = "gr-qc",
    journal = "Class. Quant. Grav.",
    volume = "42",
    pages = "3",
    year = "2025"
}

@inbook{Birrell_Davies_1982, place={Cambridge}, series={Cambridge Monographs on Mathematical Physics}, title={Quantum field theory in curved spacetime}, booktitle={Quantum Fields in Curved Space}, publisher={Cambridge University Press}, author={Birrell, N. D. and Davies, P. C. W.}, year={1982}, pages={36–88}, collection={Cambridge Monographs on Mathematical Physics}}

@article{sriramkumar_padmanabhan_2002,
  author       = {L. Sriramkumar and T. Padmanabhan},
  title        = {{Probes of the Vacuum Structure of Quantum Fields in Classical Backgrounds}},
  journal      = {International Journal of Modern Physics D},
  volume       = {11},
  number       = {1},
  pages        = {1--34},
  year         = {2002},
  doi          = {10.1142/S0218271802001354},
}

@article{Dubey:2025wws,
  author  = {Dubey, Nitesh K. and Kolekar, Sanved},
  title   = {Memory Effects and Entanglement Dynamics of Finite Time Acceleration},
  year    = {2025},
  month   = aug,
  note    = {arXiv:2508.07830 [gr-qc]}
}

@article{PhysRevD.111.065004,
  title = {Wigner distributions in Rindler spacetime and nonvacuum Minkowski states},
  author = {Dubey, Nitesh K. and Kolekar, Sanved},
  journal = {Phys. Rev. D},
  volume = {111},
  issue = {6},
  pages = {065004},
  numpages = {25},
  year = {2025},
  month = {Mar},
  publisher = {American Physical Society},
  doi = {10.1103/PhysRevD.111.065004},
  url = {https://link.aps.org/doi/10.1103/PhysRevD.111.065004}
}

@article{PhysRevA.107.L030203,
  title = {Observing single particles beyond the Rindler horizon},
  author = {Falcone, Riccardo and Conti, Claudio},
  journal = {Phys. Rev. A},
  volume = {107},
  issue = {3},
  pages = {L030203},
  numpages = {5},
  year = {2023},
  month = {Mar},
  publisher = {American Physical Society},
  doi = {10.1103/PhysRevA.107.L030203},
  url = {https://link.aps.org/doi/10.1103/PhysRevA.107.L030203}
}

@incollection{ARFKEN2013551,
title = {Chapter 12 - Further Topics in Analysis},
editor = {George B. Arfken and Hans J. Weber and Frank E. Harris},
booktitle = {Mathematical Methods for Physicists (Seventh Edition)},
publisher = {Academic Press},
edition = {Seventh Edition},
address = {Boston},
pages = {551-598},
year = {2013},
isbn = {978-0-12-384654-9},
doi = {https://doi.org/10.1016/B978-0-12-384654-9.00012-8},
url = {https://www.sciencedirect.com/science/article/pii/B9780123846549000128},
author = {George B. Arfken and Hans J. Weber and Frank E. Harris}
}

@article{Salton2015,
  title     = {Acceleration-assisted entanglement harvesting and rangefinding},
  author    = {Salton, Grant and Mann, Robert B. and Menicucci, Nicolas C.},
  journal   = {New Journal of Physics},
  volume    = {17},
  number    = {3},
  pages     = {035001},
  year      = {2015},
  month     = mar,
  publisher = {IOP Publishing},
  doi       = {10.1088/1367-2630/17/3/035001},
  url       = {https://doi.org/10.1088/1367-2630/17/3/035001}
}

@article{PhysRevD.14.870,
  title = {Notes on black-hole evaporation},
  author = {Unruh, W. G.},
  journal = {Phys. Rev. D},
  volume = {14},
  issue = {4},
  pages = {870--892},
  numpages = {0},
  year = {1976},
  month = {Aug},
  publisher = {American Physical Society},
  doi = {10.1103/PhysRevD.14.870},
  url = {https://link.aps.org/doi/10.1103/PhysRevD.14.870}
}

@article{PhysRevD.101.045017,
  title = {General relativistic quantum optics: Finite-size particle detector models in curved spacetimes},
  author = {Mart\'{\i}n-Mart\'{\i}nez, Eduardo and Perche, T. Rick and de S. L. Torres, Bruno},
  journal = {Phys. Rev. D},
  volume = {101},
  issue = {4},
  pages = {045017},
  numpages = {10},
  year = {2020},
  month = {Feb},
  publisher = {American Physical Society},
  doi = {10.1103/PhysRevD.101.045017},
  url = {https://link.aps.org/doi/10.1103/PhysRevD.101.045017}
}

@article{PhysRevD.103.025007,
  title = {Broken covariance of particle detector models in relativistic quantum information},
  author = {Mart\'{\i}n-Mart\'{\i}nez, Eduardo and Perche, T. Rick and Torres, Bruno de S. L.},
  journal = {Phys. Rev. D},
  volume = {103},
  issue = {2},
  pages = {025007},
  numpages = {14},
  year = {2021},
  month = {Jan},
  publisher = {American Physical Society},
  doi = {10.1103/PhysRevD.103.025007},
  url = {https://link.aps.org/doi/10.1103/PhysRevD.103.025007}
}

@article{PhysRevD.109.045018,
  title = {Fully relativistic entanglement harvesting},
  author = {Perche, T. Rick and Polo-G\'omez, Jos\'e and Torres, Bruno de S. L. and Mart\'{\i}n-Mart\'{\i}nez, Eduardo},
  journal = {Phys. Rev. D},
  volume = {109},
  issue = {4},
  pages = {045018},
  numpages = {17},
  year = {2024},
  month = {Feb},
  publisher = {American Physical Society},
  doi = {10.1103/PhysRevD.109.045018},
  url = {https://link.aps.org/doi/10.1103/PhysRevD.109.045018}
}

@article{PhysRevD.105.065016,
  title = {Harvesting entanglement from complex scalar and fermionic fields with linearly coupled particle detectors},
  author = {Perche, T. Rick and Lima, Caroline and Mart\'{\i}n-Mart\'{\i}nez, Eduardo},
  journal = {Phys. Rev. D},
  volume = {105},
  issue = {6},
  pages = {065016},
  numpages = {24},
  year = {2022},
  month = {Mar},
  publisher = {American Physical Society},
  doi = {10.1103/PhysRevD.105.065016},
  url = {https://link.aps.org/doi/10.1103/PhysRevD.105.065016}
}

@article{Ruep:2021fjh,
    author = "Ruep, Maximilian H.",
    title = "{Weakly coupled local particle detectors cannot harvest entanglement}",
    eprint = "2103.13400",
    archivePrefix = "arXiv",
    primaryClass = "quant-ph",
    doi = "10.1088/1361-6382/ac1b08",
    journal = "Class. Quant. Grav.",
    volume = "38",
    number = "19",
    pages = "195029",
    year = "2021"
}

@article{PhysRevD.7.2850,
  title = {Nonuniqueness of Canonical Field Quantization in Riemannian Space-Time},
  author = {Fulling, Stephen A.},
  journal = {Phys. Rev. D},
  volume = {7},
  issue = {10},
  pages = {2850--2862},
  numpages = {0},
  year = {1973},
  month = {May},
  publisher = {American Physical Society},
  doi = {10.1103/PhysRevD.7.2850},
  url = {https://link.aps.org/doi/10.1103/PhysRevD.7.2850}
}

@article{RevModPhys.80.787,
  title = {The Unruh effect and its applications},
  author = {Crispino, Lu\'{\i}s C. B. and Higuchi, Atsushi and Matsas, George E. A.},
  journal = {Rev. Mod. Phys.},
  volume = {80},
  issue = {3},
  pages = {787--838},
  numpages = {0},
  year = {2008},
  month = {Jul},
  publisher = {American Physical Society},
  doi = {10.1103/RevModPhys.80.787},
  url = {https://link.aps.org/doi/10.1103/RevModPhys.80.787}
}

@article{Hawking:1975vcx,
    author = "Hawking, S. W.",
    editor = "Gibbons, G. W. and Hawking, S. W.",
    title = "{Particle Creation by Black Holes}",
    doi = "10.1007/BF02345020",
    journal = "Commun. Math. Phys.",
    volume = "43",
    pages = "199--220",
    year = "1975",
    note = "[Erratum: Commun.Math.Phys. 46, 206 (1976)]"
}

@article{padmanabhan,
author = {PADMANABHAN, T.},
title = {THERMODYNAMICS OF HORIZONS: A COMPARISON OF SCHWARZSCHILD, RINDLER AND de SITTER SPACETIMES},
journal = {Modern Physics Letters A},
volume = {17},
number = {15n17},
pages = {923-942},
year = {2002},
doi = {10.1142/S021773230200751X},
URL = {https://doi.org/10.1142/S021773230200751X},
eprint = {https://doi.org/10.1142/S021773230200751X}
}

@article{Socolovsky:2013rga,
    author = "Socolovsky, M.",
    title = "{Rindler Space and Unruh Effect}",
    eprint = "1304.2833",
    archivePrefix = "arXiv",
    primaryClass = "gr-qc",
    month = "4",
    year = "2013"
}

@article{PhysRevD.111.045023,
  title = {Are accelerated detectors sensitive to Planck scale changes?},
  author = {Sahota, Harkirat Singh and Lochan, Kinjalk},
  journal = {Phys. Rev. D},
  volume = {111},
  issue = {4},
  pages = {045023},
  numpages = {13},
  year = {2025},
  month = {Feb},
  publisher = {American Physical Society},
  doi = {10.1103/PhysRevD.111.045023},
  url = {https://link.aps.org/doi/10.1103/PhysRevD.111.045023}
}

@article{Lin:2024ihr,
    author = "Lin, Kuan-Nan and Ievlev, Evgenii and Good, Michael R. R. and Chen, Pisin",
    title = "{Classical acceleration temperature from evaporated black hole remnants and accelerated electron-mirror radiation}",
    eprint = "2402.16137",
    archivePrefix = "arXiv",
    primaryClass = "gr-qc",
    reportNumber = "FTPI-MINN-24-07",
    doi = "10.1140/epjc/s10052-024-12991-4",
    journal = "Eur. Phys. J. C",
    volume = "84",
    number = "6",
    pages = "641",
    year = "2024"
}

@article{PhysRevD.94.065010,
  title = {Mirror reflections of a black hole},
  author = {Good, Michael R. R. and Anderson, Paul R. and Evans, Charles R.},
  journal = {Phys. Rev. D},
  volume = {94},
  issue = {6},
  pages = {065010},
  numpages = {12},
  year = {2016},
  month = {Sep},
  publisher = {American Physical Society},
  doi = {10.1103/PhysRevD.94.065010},
  url = {https://link.aps.org/doi/10.1103/PhysRevD.94.065010}
}

@book{Gradshteyn2007,
  title     = {Table of Integrals, Series, and Products},
  author    = {Gradshteyn, I. S. and Ryzhik, I. M.},
  edition   = {7th},
  year      = {2007},
  publisher = {Elsevier/Academic Press},
  address   = {Amsterdam},
}

@article{PhysRevD.102.085013,
  title = {Unruh-deWitt detectors in quantum superpositions of trajectories},
  author = {Foo, Joshua and Onoe, Sho and Zych, Magdalena},
  journal = {Phys. Rev. D},
  volume = {102},
  issue = {8},
  pages = {085013},
  numpages = {15},
  year = {2020},
  month = {Oct},
  publisher = {American Physical Society},
  doi = {10.1103/PhysRevD.102.085013},
  url = {https://link.aps.org/doi/10.1103/PhysRevD.102.085013}
}

@article{Barman:2025aqt,
    author = "Barman, Dipankar and Majhi, Bibhas Ranjan",
    title = "{Can spacetime fluctuations generate entanglement between co-moving accelerated detectors?}",
    eprint = "2504.12674",
    archivePrefix = "arXiv",
    primaryClass = "gr-qc",
    doi = "10.1016/j.physletb.2025.139631",
    journal = "Phys. Lett. B",
    volume = "868",
    pages = "139631",
    year = "2025"
}

@article{candelas1976quantum,
  author    = {P. Candelas and D. J. Raine},
  title     = {Quantum Field Theory on Incomplete Manifolds},
  journal   = {Journal of Mathematical Physics},
  volume    = {17},
  number    = {11},
  pages     = {2101--2112},
  year      = {1976},
  doi       = {10.1063/1.522850},
  publisher = {American Institute of Physics}
}

@article{Fewster:2016ewy,
    author = "Fewster, Christopher J. and Ju{\'a}rez-Aubry, Benito A. and Louko, Jorma",
    title = "{Waiting for Unruh}",
    eprint = "1605.01316",
    archivePrefix = "arXiv",
    primaryClass = "gr-qc",
    doi = "10.1088/0264-9381/33/16/165003",
    journal = "Class. Quant. Grav.",
    volume = "33",
    number = "16",
    pages = "165003",
    year = "2016"
}

\end{document}